\documentclass[twocolumn,english,prb]{revtex4-1}
\usepackage{hyperref}
\usepackage{amsmath}
\usepackage{amssymb}
\usepackage{eufrak}
\usepackage{graphicx}
\graphicspath{{Pictures/}}
\usepackage[caption=false]{subfig}

\renewcommand{\Re}{\operatorname{Re}}
\renewcommand{\Im}{\operatorname{Im}}
\newcommand{\Tr}{\operatorname{Tr}}

\begin{document}

\title{Paraconductivity of pseudogapped superconductors}

\author{Igor Poboiko $^{1,2}$ and Mikhail Feigel'man $^{3,2}$}

\affiliation{$^{1}$ Skolkovo Institute of Science and Technology, Moscow, Russia}

\affiliation{$^{2}$ National Research University "Higher School
of Economics", Moscow, Russia}

\affiliation{$^{3}$ L. D. Landau Institute for Theoretical Physics, Chernogolovka,
Moscow region, Russia}


\begin{abstract}
We calculate Aslamazov-Larkin  paraconductity $\sigma_{AL}(T)$
 for a model of  strongly disordered superconductors (dimensions $d=2,3$) with a 
large pseudogap whose magnitude strongly  exceeds transition temperature $T_c$. 
We show that, within Gaussian approximation over Cooper-pair fluctuations,
paraconductivity is just twice larger that the classical AL result at the same 
$\epsilon = (T-T_c)/T_c$.  Upon decreasing $\epsilon$,  Gaussian approximation
is violated due to local fluctuations of pairing fields that become relevant
at $\epsilon \leq \epsilon_1 \ll 1 $.  Characteristic scale $\epsilon_1 $
is  much larger than the width $\epsilon_2$ of the thermodynamical critical region,
that is determined via  the Ginzburg criterion, $\epsilon_2 \approx \epsilon_1^d$.
We argue that in the intermediate region $\epsilon_2 \leq \epsilon \leq \epsilon_1$
paraconductivity follows the same AL power law, albeit with another
(yet unknown) numerical prefactor. At further decrease of the temperature,
all kinds of fluctuational corrections become strong at 
$\epsilon \leq \epsilon_2$; in particular, conductivity occurs to be
 strongly  inhomogeneous in real space.
\end{abstract}

\maketitle

\section{Introduction}

Strongly disordered superconductors near quantum phase transition  into an insulator state
attract great interest during last years~\cite{t1,t2,t3,t4,e1,e2,e3,e4,NP2011,SacepeTiN,Sacepe2009,N1,Teun1}.
On experimental side~\cite{e1,e2,e3,e4,NP2011,SacepeTiN,Sacepe2009,N1,A1,Teun1},  new methods which became available,
like low-temperature scanning tunneling spectroscopy which makes it possible to study properties of 
superconducting state locally with a nanometer-scale resolution. As a result~\cite{NP2011,SacepeTiN} 
an existence of a strong density-of-states (DoS) suppression  at temperatures much above the superconducting transition
 $T_c$ was demonstrated. Such a phenomenon is called \textit{pseudogap}, in some rough analogy to the phenomenon known for 
under-doped high-$T_c$ oxide superconductors; however, the origin of pseudogap in usual strongly disordered
superconductors like InO$_x$, see Ref.~\onlinecite{t1,t2}, is unrelated to various  courses of pseudogap origin,
discussed  in relation to HTSC. A detailed  semi-quantitative
theory of  superconductivity, starting from BCS-like model with localized single-electron states (near 3D Anderson 
localization transition) was developed in Ref.~\onlinecite{t1,t2},  elaborating an approach proposed originally in
\cite{ML} and developed numerically in~\cite{Ghosal}.
 
One of most general phenomenon inherent to disordered superconductors is known to be  fluctuational conductivity
(paraconductivity) predicted long ago by Aslamazov and Larkin~\cite{AslamazovLarkin}.  It is due to appearance of
fluctuational (with finite life-time)  Cooper pairs at temperatures slightly above $T_c$.  Aslamazov-Larkin (AL)
paraconductivity is especially universal in 2D superconductors, where additional conductance \textit{per square}
is  
$$\sigma_{AL}^\Box = \frac{e^2}{16\hbar}\frac{T}{T-T_c}$$
independently of any microscopic parameters.  This result is usually considered to be
 valid as long as $\sigma_{AL}$ is much smaller than Drude conductance of the metal $\sigma_0$,
 i.e. at  $\epsilon \equiv T/T_c - 1 \gg \mathrm{Gi} = e^2/16\hbar\sigma_0$,
that is, in the region of Gaussian fluctuations.  In bulk systems paraconductivity is less singular,
$\sigma_{AL} \propto (T-T_c)^{-1/2}$.

More close to the transition point, within fluctuational region 
$\epsilon \leq \mathrm{Gi}$,  interaction between superconducting fluctuations become important and results in 
the universal scaling behavior of thermodynamics quantities~\cite{PokrovskiiPatashinskiiBook}, that is determined
exclusively by space dimensionality and order parameter symmetry.
In what concerns kinetic properties (like conductivity) the situation is less clear.
Ref.~\onlinecite{LO2001} provided arguments in favor of the statement that paraconductivity is more sensitive 
to nonlinear effects and deviates from classical AL form already at $\epsilon \leq \sqrt{\mathrm{Gi}}$,
that is, parametrically far from the scaling region.  Basically, the arguments of Ref.~\onlinecite{LO2001} were based
upon the suppression of the electron density of states (DoS) due to  superconducting fluctuations: reduced DoS leads to
suppression of the electron-electron inelastic rate; in turn, that leads to an increase of the order parameter 
relaxation time $\tau_{GL}$, with respect to its value known from the Gaussian approximation, 
$\tau_{GL}^{(0)} = \pi \hbar/8(T-T_c)$.  Since paraconductivity $\sigma_{AL}$ can be generally shown to be 
proportional to the product $T\tau_{GL}$, the above consideration suggests its more singular behavior due to
fluctuational suppression of the DoS. However, detailed calculations of the proposed effect were
performed~\cite{LO2001}  for the case when strong depairing is present and the whole effect  is anyway weak;
it remained unclear if indeed temperature behavior of paraconductivity changes qualitatively in the range
$\epsilon \leq \sqrt{\mathrm{Gi}}$.

In the present paper we provide an analysis of similar problem from a different perspective. Namely, we consider
very strongly disordered superconductor with a well-developed pseudogap $\Delta_P$.   
An existence of pseudogap $\Delta_P$ is due to i) localized nature of single-electron eigenstates $\psi_i(\boldsymbol{r})$,
 and ii) phonon-induced attraction between electrons
which leads to formation of localized electron pairs (with opposite spins) populating eigenstates $\psi_i(\boldsymbol{r})$.
The energy gain due to formation of such a pair is $\Delta_P$.  Next, hybridization matrix elements $J_{ij}$
provide virtual hopping of electron pairs between different localized eigenstates. If this hopping is sufficiently
strong, a superconducting coherent state is formed below some critical temperature $T_c$; \, 
(for detailed theory of pseudogaped superconductivity see~\cite{t2}).

Below we consider the case of $\Delta_P$ that is much larger than $T_c$, like it was found
in InO$_x$ thick films studied in Ref.\onlinecite{NP2011}. In such a case one may neglect, to a first approximation over 
$T_c/\Delta_P \ll 1$, the presence of single-electron states:  the single-particle DoS will be set to zero.
We will show, nevertheless, that the whole qualitative picture of critical fluctuations, including their
dynamics, remains the same as for usual disordered superconductors, as long as we stick to the Gaussian fluctuation
region.  The major difference we found is that now $\tau_{GL}^{(0)} = \pi \hbar/4(T-T_c)$, i.e. twice larger
 than the result of the standard theory.   

This result is valid as long as thermal fluctuations are weak
and their interaction can be neglected.  For such a region to exist at $\epsilon \leq 1$ in a pseudogapped 
superconductor, a special  assumption is necessary; namely, we consider the model with interaction matrix
elements $J_{ij}$ possessing  large coordination number $Z \gg 1$ for the relevant eigenstates 
which  have eigenenergies $\varepsilon_i ,\varepsilon_j$ located within about $T_c$ from Fermi energy.
The presence of large parameter $Z$ allows us to derive a dynamical Ginzburg-Landau functional for superconducting
fluctuations at $T$ near $T_c$ and to calculate paraconductivity at $\epsilon \geq \epsilon_1 \ll 1$ where explicit value of
$\epsilon_1$ depends both on $Z$ and on space dimensionality $d$ (we consider $d=2,3$ ).
At smaller $\epsilon$ interaction between fluctuations becomes strong enough to  affect kinetic coefficient of the
Ginzburg-Landau functional, thus the kinetic problem cannot be solved
analytically; however,  we provide some arguments in favor of the same type power-law singularity
in paraconductivity  $\sigma_{AL} \propto (T-T_c)^{(d-4)/2}$
to exist down to much smaller values of  $\epsilon \geq \epsilon_2$.
Here  $\epsilon_2$ provides a boundary of the region where all thermodynamic fluctuational effects become strong,
it is analogous to the Ginzburg parameter in the usual theory of second-order phase transitions; the important point
is that  $\epsilon_2 \ll \epsilon_1$ as long as $\epsilon_1 \ll 1$.

The rest of the paper is organized as follows: in Sec.\ref{sec:Model} we formulate our model based upon Anderson pseudospin~
\cite{AndersonPseudo} representation of the even-only sector of BCS Hamiltonian for localized single-electron states; we
provide initial mean-field-like analysis in \ref{sec:MeanFieldStatic} and then in  Sec.\ref{sec:KeldyshTechnique} 
 we develop Popov-Fedotov semionic diagrammatic technique that is convenient to treat long-range and long-time 
properties of the model near the critical
point.  Sec. \ref{sec:Gaussian} is devoted to the derivation of the dynamic Ginzburg-Landau functional and to the calculation of the paraconductivity within Gaussian approximation for 2D and 3D systems. In Sec. \ref{sec:LocalNoise} we analyze leading non-Gaussian
corrections and estimate characteristic temperature scale $\epsilon_1$; we find that it scales as $Z^{-1/2}$ and
$Z^{-2/3}$ in 2D and 3D cases, correspondingly; we also analyze the effect of these non-Gaussian corrections upon
dependence of $\sigma_{AL}$ on $\epsilon$. Then, in Sec.\ref{sec:RandomFieldCorrections} we consider all other effects beyond the leading
Gaussian approximation; these effects are:  a) the lack of self-averaging due to strong spatial fluctuations 
of disorder, and b) infra-red dominated thermal fluctuations of collective modes. We show that corresponding
reduced temperature scale $\epsilon_2 \propto 1/Z$ in 2D model, and $\propto 1/Z^2$ in 3D; note that it is the
same scaling as it is known for  the Ginzburg number $\mathrm{Gi}$ in usual phase transition theory.
 Sec. \ref{sec:Conclusions} contains our conclusions.
Some technical details are presented in Appendices \ref{sec:AppendixSCBA},
\ref{sec:AppendixKeldysh}, \ref{sec:ImpurityTechnique} and \ref{sec:AppendixFluctuations}.

\section{The model and diagram technique}
\label{sec:Model}

The starting point of our approach is representation of the \textit{paired} electron system in terms of
pseudospin operators introduced long ago by P.W.Anderson~\cite{AndersonPseudo}:
\begin{equation}
S_i^- = a_{i\downarrow}a_{i\uparrow} \quad S_i^+ = a_{i\uparrow}^\dagger a_{i\downarrow}^\dagger  
\quad 2S_i^z = 1 - a_{i\uparrow}^\dagger a_{i\uparrow} -  a_{i\downarrow}^\dagger a_{i\downarrow}
\label{spins}
\end{equation}
Here operators $a_{i\uparrow}$ and $a_{i\downarrow}$, and, correspondingly,
$a_{i\uparrow}^+$ and $a_{i\downarrow}^+$ represent electron annihilation (creation) operators
for $i$-th single-particle eigenstate $\psi_i(\boldsymbol{r})$ which are assumed to be localized.
  Then operators $S_i^\alpha$ introduced in (\ref{spins}) obey standard
 spin-$1/2$ commutation relations. The Hilbert space spanned by the set of $S_i^\alpha$ operators
constitutes a part of the whole Hilbert space of the electron system; namely, we omit the states with
some eigenstates $\psi_i(\boldsymbol{r})$ to be single-occupied. This is reasonable approximation as long as
two-electron local binding energy $\Delta_P$ is much larger than all energy/temperature scales relevant
for the problem to be considered, see Ref.~\cite{t2}. 

The minimal Hamiltonian that describes development of superconducting correlations between localized
electron pairs is of the form
\begin{equation}
\label{eq:PseudospinHamiltonian}
H=-2\sum_{i}\varepsilon_{i}S_{i}^{z}-\frac{1}{2}\sum_{ij}J_{ij}(S_{i}^{+}S_{j}^{-}+h.c),
\end{equation}
where $\varepsilon_i$ are single-electron eigenvalues which
 are assumed to be distributed independently with the box 
distribution function $P(\varepsilon) = \frac{1}{2W}\theta(W-|\varepsilon|)$. The exact shape of the distribution function is important only for the $T_c$ definition; as we will show below, the critical behavior near the transition (such as paraconductivity) depend only on the shape of the distribution function at $\varepsilon \lesssim |T - T_c|$. As long as density of states $\nu_0 = P(0)$ is finite, all the results will hold the same.

Matrix elements 
$J_{ij}\equiv J(\boldsymbol{r}_{i}-\boldsymbol{r}_{j}) \propto \int d^d r \psi_i^2(\boldsymbol{r}) \psi_j^2(\boldsymbol{r})$ 
actually depend in nontrivial way on the distance $\boldsymbol{r} = \boldsymbol{r}_{i}-\boldsymbol{r}_{j} $
as well as on the energy difference $\varepsilon_i-\varepsilon_j$; to simplify the problem, we  employ below a model
where $J_{ij}$
are assumed to have large radius $R\gg1$, and its Fourier transform  takes the form $J(\boldsymbol{p}) = J(1 - p^2 R^2)$
in the long-wavelength limit.

The disorder is assumed to be large, and the temperature is assumed to be small, so that $W \gg J \gg T$.

\subsection{Mean field critical temperature and order parameter}
\label{sec:MeanFieldStatic}
The BCS order parameter, which is the anomalous average, corresponds to non-zero in-plane spin magnetization $\left\langle S_{i}^{x,y}\right\rangle$. The natural choice for the order parameter for the mean-field treatment is thus the following:
\begin{equation}
\label{eq:OrderParameter}
\Phi_{i}^{\alpha}=\sum_{j}J_{ij}\left\langle S_{j}^{\alpha}\right\rangle,\quad \alpha = x,y,
\end{equation}
while the ordinary superconducting complex order parameter takes the form $\Delta = \Phi^{x} + i \Phi^{y}$.

In the mean-field approximation one decouples spins living in effective magnetic field created collectively by other spins:
\begin{equation}
\label{eq:StaticMeanField} H_{MF}=-\sum_{i,\alpha}h_{i}^{\alpha}\sigma_{i}^{\alpha},
\end{equation}
with Pauli matrix $\sigma_i^\alpha = 2 S_i^\alpha$ and effective magnetic field $\boldsymbol{h}_{i}=(\Phi_{i}^{x},\Phi_{i}^{y},\varepsilon_{i})$. The Hamiltonian \eqref{eq:StaticMeanField} yields trivial partition function for each spin $Z_i=2\cosh(\beta|h_{i}|)$. From this partition function we immediately extract the average magnetization $\left\langle S_{i}^{\alpha}\right\rangle =\frac{T}{2}\frac{\partial\ln Z_{i}}{\partial h_{i}^{\alpha}}$, which yields following self-consistency equations:
\begin{equation}
\sum_{j}J_{ij}\eta_{j}\Phi_{j}^{\alpha}=\Phi_{i}^{\alpha},\quad \eta_j = \frac{\tanh\beta\sqrt{\varepsilon_{j}^{2}+\Phi_{j}^{2}}}{2\sqrt{\varepsilon_{j}^{2}+\Phi_{j}^{2}}}.
\end{equation}
Note that matrix $J_{ij} \eta_j$ entering these equations  is non-Hermitian; however, it can be made Hermitian trivially by rescaling $\Phi_i \mapsto \Phi_i / \sqrt{\eta_i}$ yielding a new matrix $\sqrt{\eta_i \eta_j} J_{ij}$. These equations acquire a non-trivial solution if the matrix has a unity eigenvalue. The critical temperature $T_c$ can be defined as the highest temperature that
is consistent with the same condition for $\boldsymbol{\Phi} = 0$.

Under the assumption of a very large interaction radius $R$, one can simply average all $\eta_i$ over $\varepsilon_i$ and assume a homogeneous order parameter $\Phi_i^\alpha \equiv \Phi^\alpha$. These simplifications lead to the self-consistency equation, which is nearly equivalent to the BCS one:
\begin{equation}
1=\frac{J}{2}\int d\varepsilon P(\varepsilon)\frac{\tanh\left(\beta\sqrt{\varepsilon^{2}+\boldsymbol{\Phi}^{2}}\right)}{\sqrt{\varepsilon^{2}+\boldsymbol{\Phi}^{2}}}.
\label{TcEq}
\end{equation}
For the box-shaped distribution $P(\varepsilon)$, the critical temperature that follows from this equation reads:
\begin{equation}
\label{eq:TcMeanField}
T_c = \frac{4e^{\gamma}}{\pi}We^{-1/g},\quad g = \nu_0 J
\end{equation}
Here $\gamma \approx 0.577$ is Euler's constant. The parametric dependence $T_c \sim W e^{-1/g}$ is not sensitive to the exact shape of the distribution function, only the numerical prefactor is.  Relevant parameters of the distribution function  are:  non-zero DoS $\nu_0 = P(0)$ and a typical width $W$.

Note the absence of factor 2 in denomenator in the argument of
$\tanh $ in Eq.\eqref{TcEq}; this is due to
the absence of odd-electron states in the Hilbert space of our model.  In result, the value of $T_c$ is twice larger
than in the BCS theory.

The value of the order parameter at $T=0$ is given exactly by the standard
BCS formula
\begin{equation}
\label{eq:Delta}
\Phi(0) =  2We^{-1/g}
\end{equation}

Eq.\eqref{TcEq} is exactly valid in the limit of $R\to \infty$ only. For a finite $R$, in order to find $T_c$,
one should consider actual matrix
$J_{ij} \sqrt{\eta_i \eta_j}$ for a given realization of $\{\varepsilon_i\}$ and look for $T = \beta^{-1}$ that
 corresponds to its highest eigenvalue equal to 1.
Below we provide major results of the corresponding numerical analysis (the details are presented in the Appendix \ref{sec:AppendixSCBA}). 

\begin{figure}
	\includegraphics[width=\columnwidth]{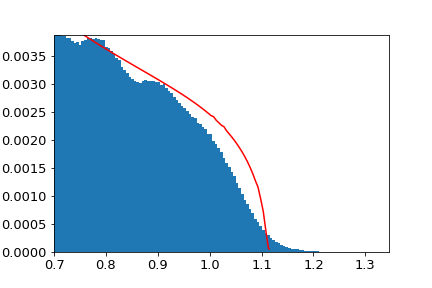}
	\caption{DoS of the random matrix $J_{ij} \sqrt{\eta_i \eta_j}$ for $200 \times 200$ system with following parameters: $R = 5$, $J = 1$, $W = 3$, $\beta \approx \beta_c \approx 60$. Red curve: analytical fit given by SCBA approximation, see Appendix \ref{sec:AppendixSCBA} for more details.}
	\label{fig:tail1}
\end{figure}

Taking into account the large yet finite $R$ leads to the shift of the criterion for superconductivity from the value given by \eqref{eq:TcMeanField} to the larger temperatures. Typical behavior of the DoS at large $R$ obtained by means of numerical diagonalization and disorder averaging is shown on Fig. \ref{fig:DOS}. The ``main body'' of the DoS fits reasonably well with the one predicted by the SCBA-approximation developed in the Appendix, and the width of the exponential tail is proportional to the Ginzburg number, $\mathrm{Gi} \sim \rho^{2 / (4 - d)}$ with $\rho$ given by \eqref{eq:rho2D} and \eqref{eq:rho3D}. Oscillatory behavior is an artifact caused by finite size of the system.

\subsection{Semionic description and Keldysh diagram technique}
\label{sec:KeldyshTechnique}
In order to study the dynamical properties of the order parameter and develop a diagram technique, we choose the Fedotov-Popov representation for spin-$\frac{1}{2}$ operators\cite{FedotovPopov,Shtyk}. Namely, for each spin we introduce a two-component spinor $\psi = (\psi_{\uparrow}, \psi_{\downarrow})$ describing a pair of fermions (called semions for the reason that will become clear soon), and represent spin operators in terms of semions (below $\hat{\sigma}^\alpha$ is the set of Pauli matrices
acting in the $({\uparrow},{\downarrow})$ space):
\begin{equation}
S^\alpha_{i} = \frac{1}{2} \psi^\dagger_{i}\hat{\sigma}^\alpha \psi_i.
\end{equation}
The physical subspace contains 2 states and corresponds to the presence of exactly one  semion: 
$\psi^\dagger_\alpha \psi_\alpha = 1$; in order to get rid of two extra (unphysical)  degrees of freedom, one should  introduce an imaginary chemical potential $\mu = -\frac{i}{2} \pi T$  for the semions~\cite{FedotovPopov}. In the imaginary-time Matsubara representation, such an addition to the chemical potential is equivalent to the  additional phase shift equal to $\pm \pi/2$ for
fermionic fields translation over period along the imaginary time axis:
 $\psi_\alpha (\tau + \beta) =  ( \pm i) \psi_\alpha (\tau)$, thus these modified fermions were coined "semions".

 The Hamiltonian \eqref{eq:PseudospinHamiltonian} expressed in term of semionic degrees of freedom reads:
\begin{equation}
\label{eq:SemionHamitlonian}
H=-\sum_{i}\varepsilon_{i}\psi_{i}^{\dagger}\hat{\sigma}^{z}\psi_{i}-\frac{1}{4}\sum_{ij,\alpha}(\psi_{i}^{\dagger}\hat{\sigma}^{\alpha}\psi_{i})J_{ij}(\psi_{j}^{\dagger}\hat{\sigma}^{\alpha}\psi_{j}).
\end{equation}

This expression for the Hamiltonian allows us to build a Keldysh diagram technique for calculation of spin-spin correlation functions. After decoupling the four semion interaction using the Hubbard-Stratanovich order parameter field $\Phi$ (see Appendix \ref{sec:AppendixKeldysh} for the detailed derivation), we arrive at the following Keldysh action describing semionic as well as order parameter degrees of freedom:
\begin{multline}
\label{eq:KeldyshAction}
iS[\bar{\psi},\psi,\Phi]=i\int dt\Big(-\Phi^{\alpha}\hat{J}^{-1}\check{\tau}_{x}\Phi^{\alpha}+\\
+\bar{\psi}\left(\hat{G}^{-1} + \frac{1}{\sqrt{2}}\check{\Gamma}_{\mu}\hat{\sigma}^{\alpha}\Phi_{\mu}^{\alpha}\right)\psi\Big)
\end{multline}
Here index $\mu \in \{cl, q\}$ denotes ``classical'' and ``quantum'' Keldysh components; vertices $\check{\Gamma}_{cl} = \check{\tau}_0$, $\check{\Gamma}_{q} = \check{\tau}_x$ with $\check{\tau}_\alpha$ being Pauli matrices acting in Keldysh space; and $\hat{G}^{-1} = i \partial_t + \varepsilon_i \hat{\sigma}^z$ is a matrix, diagonal in the real space.

Quadratic part of the action is used to build the following ``bare'' propagators for the order parameter
 $L^{\alpha \beta}(t-t^\prime) = i \left\langle \Phi^\alpha(t) \Phi^\beta(t^\prime)\right\rangle$ 
(which appears to be diagonal in
 spin space $(L^{(0)})^{\alpha \beta}  = \delta^{\alpha \beta} L^{(0)}$):
\begin{equation}
\label{eq:OrderBarePropagator}
L_{R/A}^{(0)}(\omega, \boldsymbol{q}) = J(\boldsymbol{q})/2,
\end{equation}
and for the semions $G_{\sigma \sigma^\prime}(t-t^\prime) = -i \left\langle\psi_\sigma(t) \psi^\dagger_{\sigma^\prime}(t^\prime)\right\rangle$ (with $\sigma,\sigma^\prime \in \{\uparrow, \downarrow\}$):
\begin{multline}
\label{eq:SemionBarePropagator}
\hat{G}^{(0)}_{R/A}(\omega)=\begin{pmatrix}(\omega\pm i\gamma+\varepsilon)^{-1} & 0\\
0 & (\omega\pm i\gamma-\varepsilon)^{-1}
\end{pmatrix}=\\
=\hat{\mathbb{P}}^{\uparrow}G_{R/A}^{\uparrow}(\omega)+\hat{\mathbb{P}}^{\downarrow}G_{R/A}^{\downarrow}(\omega).
\end{multline}
Here $\hat{\mathbb{P}}^{\uparrow,\downarrow} = \frac{1}{2} (1 \pm \hat{\sigma}^z)$ are the projectors onto $z$ axis.  Imaginary part $\gamma$ should be taken positive infinitesimal.

Finally, in the equilibrium, the standard Keldysh relation holds:
\begin{equation}
L_K(\omega) = \mathfrak{B}(\omega) \Delta L(\omega),\quad 
\mathfrak{B}(\omega)=\coth\frac{\beta\omega}{2},
\end{equation}
 and
\begin{multline}
\label{eq:SemionEquilibriumRelation}
G_K(\omega) = \mathfrak{F}(\omega) \Delta G(\omega),\quad \mathrm{with} \\
\mathfrak{F}(\omega) = \mathfrak{f}(\omega) - \frac{i}{\cosh \frac{\beta\omega}{2}},\quad \mathfrak{f}(\omega) = \tanh \beta \omega,
\end{multline}
where the shorthand notation $\Delta(\dots) = (\dots)_R - (\dots)_A$ is introduced. The only modification is that semions acquire an imaginary part in their distribution function, which is due to the imaginary chemical potential $\mu = -i \pi T / 2$. That does  not produce any problem since semions
themselves do not correspond to any physical degrees of freedom, while original spins do.

Below we will use the developed diagram technique in order to calculate the order parameter correlation function 
$L(\omega, \boldsymbol{q})$ above the transition temperature, but in its close vicinity, where critical slowing down takes place.

\subsection{Electric current}
Anderson pseudospin operators $S^{\pm}_{i}$ create and annihilate pair of electrons on site $i$. The electromagnetic gauge transformation thus acts as $U(1)$ rotation on the spin operators $S^{\pm}_i \mapsto S^{\pm}_i e^{\pm2 i e \alpha(\boldsymbol{r}_i)}$ (with $e$ being electron charge, while speed of light is taken $c = 1$). Accompanied by the gauge transformation for the vector potential $\boldsymbol{A}(\boldsymbol{r}) \mapsto \boldsymbol{A}(\boldsymbol{r}) - \nabla \alpha$ this should leave the action \eqref{eq:KeldyshAction} unchanged.

Real space enters problem via the $\hat{J} = J(\hat{p} = -i \nabla)$ matrix. The gauge field $\boldsymbol{A}$ thus enters the action by replacing momentum by the ``covariant derivative'' $\hat{\boldsymbol{P}}=\hat{\boldsymbol{p}}-2e\boldsymbol{A}\hat{\sigma}^{y}$. The long-wavelength limit corresponds to $\hat{J}^{-1} \equiv \hat{J}^{-1}(\hat{\boldsymbol{P}})\approx J^{-1} (1 + \hat{\boldsymbol{P}}^2 R^2)$.

The electrical current induced by Cooper pairs can be extracted from the action using the following relation:
\begin{equation}
\label{eq:ElectricCurrent}
\boldsymbol{j}=\frac{\delta S}{\delta\boldsymbol{A}}=\frac{4 e R^2}{J}\Phi^{\alpha}\left[\hat{\sigma}^{y}_{\alpha \beta}\hat{\boldsymbol{p}}-2e\boldsymbol{A}\delta_{\alpha \beta}\right]\Phi^{\beta}.
\end{equation}
This relation holds on the classical field theory level, and it is translated to an operator identity of the corresponding quantum  theory.

\section{Gaussian fluctuations and paraconductivity}
\label{sec:Gaussian}

In this Section we consider the fluctuation propagator  of the order parameter  $L(\omega, \boldsymbol{q})$
in the simplest Gaussian approximation, and calculate the corresponding fluctuation contribution to  
electric conductivity.

\subsection{Order parameter propagator}
\label{sec:OrderParameterMeanField}

\begin{figure}
	\centering
	\includegraphics[width=0.8\columnwidth]{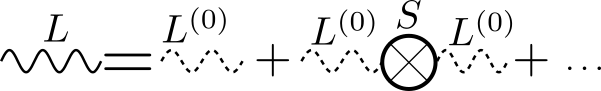}
	\includegraphics[width=0.8\columnwidth]{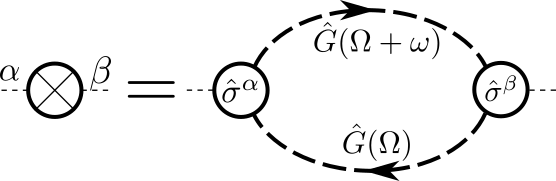}
	\caption{Upper figure: Dyson series for the order parameter Green function $L$ given by Eq. \eqref{eq:DysonEquation}; dashed wavy line corresponds to $\hat{L}^{(0)} = \hat{J}/2$. The crossed circle, shown in the lower figure, is spin-spin correlation function $\hat{S}$ given by \eqref{eq:OrderParameterSelfEnergy} acts as the ``polarization operator'' for the order parameter propagator. The crossed circle indicates that it is a diagonal operator in real space (but it still has non-trivial time structure). Spin-space indices $\alpha,\beta \in \{x,y,z\}$; vertices have also Keldysh structure $\check{\Gamma}_\mu$, $\mu \in \{cl, q\}$. Dashed lines correspond to the semionic Green functions given by \eqref{eq:SemionBarePropagator}}
	\label{fig:OrderParameterSelfEnergy}
\end{figure}

On the Gaussian level, the order parameter Green function $\check{L}$ is given by the Dyson series shown on Fig. \ref{fig:OrderParameterSelfEnergy}, with analytic expression given by
\begin{equation}
\label{eq:DysonEquation}
\check{L}^{-1}=(\check{L}^{(0)})^{-1}-\check{S},
\end{equation}

\begin{equation}
\label{eq:OrderParameterSelfEnergy}
S_{\mu\nu}^{\alpha\beta}(\omega)=\frac{i}{2}\int\frac{d\Omega}{2 \pi}\Tr(\check{\Gamma}_{\mu}\hat{\sigma}^{\alpha}\hat{G}(\Omega + \omega)\check{\Gamma}_{\nu}\hat{\sigma}^{\beta}\hat{G}(\Omega))
\end{equation}
The expression for the self-energy part coincides with the unperturbed spin-spin correlation function 
$S^{\alpha \beta}_i(t-t^\prime) = i \left\langle\hat{\sigma}^\alpha_i (t) \hat{\sigma}^\beta_i(t^\prime)\right\rangle$. 
Note that $\hat{S}$ is  a diagonal in real-space matrix, which depends on the on-site random energy $\varepsilon_i$.
 Since we are interested in $\left\langle L\right\rangle_{\varepsilon}$, we need to average the whole Dyson series (Fig. \ref{fig:OrderParameterSelfEnergy}). 
We employ an approximation of large radius $R$ which guarantees that propagator $L$ changes considerably
 on a long spatial scale which includes many individual "spins" $S_i$; thus we  can build  a kind of  "impurity diagram technique" with regard to random local fields $\varepsilon_i$.


The Dyson equation for the average propagator $ \left\langle \check{L}\right\rangle$ reads
\begin{equation}
\label{eq:DysonDisorder}
\left\langle \check{L}\right\rangle_{\varepsilon} ^{-1}=(\check{L}^{(0)})^{-1}-\check{\Pi}.
\end{equation}
To the leading order in large $R$ we can average all the $\check{S}_i$ independently and put $\check{\Pi} \approx \left\langle \check{S}\right\rangle_{\varepsilon}$. 
Below in Sec. \ref{sec:RandomFieldCorrections} we will take into account additional terms
beyond this simplest approximation. 

We now proceed with the calculation of the self-energy \eqref{eq:OrderParameterSelfEnergy}. The Keldysh space can be traced out immediately; the retarded ($\mu = q$, $\nu = cl$) component reads as follows:
\begin{multline}
\label{eq:SpinCorrelationFunction}
S_{R}^{\alpha\beta}(\omega)=\frac{i}{2}\int \frac{d\Omega}{2\pi}\Tr\Big(\hat{\sigma}^{\alpha}\hat{G}_{R}(\Omega+\omega)\hat{\sigma}^{\beta}\hat{G}_{K}(\Omega)+ \\
+\hat{\sigma}^{\alpha}\hat{G}_{K}(\Omega+\omega)\hat{\sigma}^{\beta}\hat{G}_{A}(\Omega)\Big).
\end{multline}
We perform all the calculations by keeping $\gamma$ finite, as we will refer to them later in Sec. \ref{sec:OrderParameterCorrection}; however, within  Gaussian approximation for fluctuations
the limit $\gamma \to 0$ is sufficient.
The terms arising after substitution of semionic bare propagators given by Eq. \eqref{eq:SemionBarePropagator} can be divided onto two groups. First group corresponds to semions residing on the same branch, $\propto G^{\uparrow}G^{\uparrow}$ or $\propto G^{\downarrow} G^{\downarrow}$. It appears to vanish in the limit $\gamma \to 0$, while for finite $\gamma$ it is odd in $\varepsilon\mapsto -\varepsilon$ and thus vanishes upon further averaging over $\varepsilon$. 
The second group, where semions residing in different branches, can itself be naturally divided into diagonal and off-diagonal in spin space parts. Introducing the unit vector in the $z$ direction $\boldsymbol{n} = (0, 0, 1)$, and performing the energy integration, we obtain following results:
\begin{equation}
\label{eq:SRstructure}
S_{R}^{\alpha\beta}(\omega) = (\delta^{\alpha\beta}-n^{\alpha}n^{\beta}) S_{R}^{(diag)}(\omega) +i\epsilon^{\alpha\beta\mu}n^{\mu} S_{R}^{(off)}(\omega),
\end{equation}
where in the limit $\gamma \ll T,\varepsilon$ we find
\begin{align}
\label{eq:SRdiag}
S_{R}^{(diag)}(\omega) &\approx \frac{\mathfrak{f}(\varepsilon) \varepsilon}{\varepsilon^2-(\omega/2 + i\gamma)^2}\\
\label{eq:SRoffdiag}
S_{R}^{(off)}(\omega) &\approx  \frac{\mathfrak{f}(\varepsilon) \omega / 2}{\varepsilon^{2}-(\omega/2+i\gamma)^{2}},
\end{align}
and $\mathfrak{f}(\varepsilon)$ is given by Eq. \eqref{eq:SemionEquilibriumRelation}. In the limit $\gamma \to +0$ these correlation functions describe trivial dynamics of a single spin precession in a constant magnetic field $\varepsilon\boldsymbol{n}$.

Next step is to perform averaging over $\varepsilon$ to calculate  $\Pi_R(\omega) \approx \left\langle S_R(\omega)\right\rangle_{\varepsilon}$. 
The off-diagonal part is odd in $\varepsilon \mapsto -\varepsilon$ even at finite $\gamma$ and vanishes upon averaging, thus the only non-trivial contribution is due to $S_R^{(diag)}(\omega)$.  In the limit $\omega \ll T$, it is natural to consider real and imaginary part of the correlation function independently. The real part is static, it  determines the critical temperature of the transition, while the imaginary part is $\omega$-dependent and describes  purely dissipative dynamics of the order parameter fluctuations:
\begin{align}
\left\langle \Re S_{R}^{(diag)}(\omega)\right\rangle &
\approx\left\langle \frac{\mathfrak{f}(\varepsilon)}{\varepsilon}\right\rangle_{\varepsilon} 
=\frac{1}{W}\ln\frac{4e^{\gamma}W}{\pi T}\\
\left\langle \Im S_{R}^{(diag)}(\omega)\right\rangle &=\pi \nu_0 \mathfrak{f}\left(\frac{\omega}{2}\right)\approx\frac{\pi\omega}{4WT}
\end{align}
Note that the major contribution to the static part comes from logarithmically broad energy range  between $T \ll \varepsilon \ll W$, while the imaginary part is given by $\varepsilon \sim \omega$ as it describes  real resonant spin-flip processes which lead to the dissipation of the order parameter fluctuations. The presence of a linear in $\omega$ term is thus a direct consequence of the non-zero single-spin density of states $\nu_0 = P(\varepsilon \ll T)$.

The above calculation leads to the following form of the order parameter propagator:
\begin{equation}
\label{eq:OrderParameterPropagator}
L_{R}(\omega,\boldsymbol{q}) = \frac{1 / 2 \nu_0}{\epsilon + q^2 \xi_0^2 - i\omega\tau},
\end{equation}
with
\begin{equation}
\label{GLparameters}
\epsilon = \ln\frac{T}{T_c}\approx \frac{T - T_c}{T_c} \ll 1, \quad \xi_0 = \frac{R}{\sqrt{g}},\quad \tau = \frac{\pi}{4 T},
\end{equation}
and $T_c$ given by the same expression as given above \eqref{eq:TcMeanField}. 
The dimensionless parameter $\epsilon$ describes the distance to the superconducting transition, $\xi_0$ corresponds to the ``zero-temperature'' coherence length, and $\tau^{-1}$ defines  the decay rate of the collective mode far from $T_c$.  At small $\epsilon$, coherence length  and 
relaxation time diverge as $\xi(\epsilon) = \xi_0 / \sqrt{\epsilon}$ and $ \tau/\epsilon $, correspondingly.

We should emphasize that the form of the propagator \eqref{eq:OrderParameterPropagator} is independent of the exact shape of the distribution function provided it has non-zero DoS $\nu_0 = P(\varepsilon = 0)$ and does not change significantly at $\varepsilon \lesssim \omega$. The only parameter sensitive to the exact shape is $T_c$.

This form of the propagator is reminiscent of the ordinary time-dependent Ginzburg-Landau (TDGL)\cite{VarlamovLarkin} theory describing the dynamics of the order parameter in the metals close to the superconducting transition. 
The difference is that in our theory $\xi_0$ does not scale with $T_c$ as it does in disordered metals, 
where $\xi_0 \sim \sqrt{D / T_c}$;
another important difference is that the  parameter $\tau$ we found if  twice larger compared to the value 
known for disordered metals, where $\tau = \pi / 8 T$.

\subsection{Fluctuational conductivity}
\label{sec:MeanFieldConductivity}

We found in the previous Section that dynamics of our order parameter appears to be similar to the usual TDGL.  Paraconductivity in superconductors above $T_c$ was calculated long time ago by Aslamazov and Larkin \cite{AslamazovLarkin}, while its calculation using TDGL formalism can be found in Ref. \cite{VarlamovLarkin}. In this Section we briefly recapitulate  the calculation and discuss the obtained results. 

In order to obtain the expression for the electric conductivity of the system, one can apply Kubo formula,
\begin{equation}
\label{sigma}
\sigma^{i j}(\omega,\boldsymbol{q})=i\frac{Q_{R}^{i j}(\omega,\boldsymbol{q})-Q_{R}^{i j}(0,\boldsymbol{q})}{\omega}.
\end{equation}
where current-current correlation function in real space-time reads
\begin{equation}
\label{eq:Qkernel}
Q_{\mu \nu}^{i j}(\boldsymbol{r}-\boldsymbol{r}^{\prime},t-t^{\prime})=-i\left\langle j_{\mu}^{i}(\boldsymbol{r},t)j_{\nu}^{j}(\boldsymbol{r}^{\prime},t^{\prime})\right\rangle,
\end{equation}
and  Eq.\eqref{sigma} contains its Fourier transform to $(\boldsymbol{q},\omega)$ representation.

\begin{figure}
	\centering
	\includegraphics[width=0.8\columnwidth]{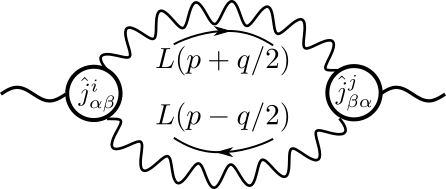}
	\caption{Diagrammatic representation of $Q$-kernel given by Eq. \eqref{eq:Qkernel}. Wavy lines correspond to the order parameter Green functions $L(\omega, \boldsymbol{q})$, and current vertices are $\hat{j}^{i}_{\alpha \beta} = \frac{4 R^2}{J} \hat{\sigma^y}_{\alpha \beta} (-i \nabla^i)$, see Eq. \eqref{eq:ElectricCurrent}}
	\label{fig:Qkernel}
\end{figure}
Within the Gaussian approximation over fluctuations,
the only diagram contributing to the $Q$-kernel is given by Fig. \ref{fig:Qkernel},
 and the corresponding expression yields (here $p_{\pm} = p \pm \frac{q}{2}$, $p = (\Omega, \boldsymbol{p})$ 
and $q = (\omega, \boldsymbol{q})$:
\begin{multline}
\label{eq:QkernelLoop}
Q_{R}^{i j}(\omega,\boldsymbol{q})=32i \nu_0^2 \xi_0^2\int\frac{d\Omega}{2\pi}\frac{d^{d}\boldsymbol{p}}{(2\pi)^{d}}p^i p^j \times\\
\times\left(L_{R}(p_{+})L_{K}(p_{-})+L_{K}(p_{+})L_{A}(p_{-})\right)
\end{multline}
In the static limit $\omega \to 0$, the expression for the uniform ($q=0$) Aslamazov-Larkin conductivity is diagonal and reads $\sigma_{AL} = i \partial Q_R / \partial \omega$, which can be further simplified:
\begin{equation}
\label{eq:Conductivity}
\sigma_{AL} = \frac{16}{d} \nu_0^2 \xi_{0}^{4}\int\frac{d\Omega}{2\pi}\frac{d^d\boldsymbol{p}}{(2\pi)^d}\boldsymbol{p}^{2}\mathfrak{B}^{\prime}(\Omega)(\Delta L(\Omega, \boldsymbol{p}))^{2}
\end{equation}
Now we substitute Eq.\eqref{eq:OrderParameterPropagator} for the propagator $L(\omega,\boldsymbol{q})$, perform integration over energy using residues and switch to integration over dimensionless momentum $P = p \xi_0 / \sqrt{\epsilon}$, to arrive at:
\begin{equation}
\label{eq:ALArbitraryD}
\sigma_{AL} = \frac{1}{\xi_{0}^{d-2}\epsilon^{2-d/2}}\frac{8T\tau}{d}\int\frac{d^{d}\boldsymbol{P}}{(2\pi)^{d}}\frac{P^{2}}{(1+P^{2})^{3}} \equiv \frac{\sigma_d}{\xi_0^{d-2} \epsilon^{2-d/2}},
\end{equation}
where $\sigma_2 = \frac{1}{8}$ and $\sigma_3 = \frac{1}{16}$.  Finally, we find paraconductivity in the form
\begin{equation}
\label{eq:AslamazovLarkin}
\sigma_{AL}=\begin{cases}
1 / 8\epsilon, & (2D)\\
1 / 16\xi_{0}\sqrt{\epsilon}, & (3D)
\end{cases}
\end{equation}
This result appears to be twice larger compared to the ordinary Aslamazov-Larkin result\cite{AslamazovLarkin}. The discrepancy can be traced back to the fact that  $\tau$ is  twice larger compared to the ordinary metals, which we have briefly mentioned above.
While  in ordinary superconductors the AL paraconductivity provides a
relatively small correction to the standard Drude conductivity $\sigma_D$, 
in our system with a large pseudogap, paraconductivity $\sigma_{AL}$  may occur to be the
dominant contribution: the only alternative conduction channel is due to  individual  electrons  hopping between 
localized states, those contribution is suppressed additionally due to $T_c \ll \Delta_P$ condition.


Below we will  study different kinds of corrections to the Gaussian approximation we used, 
and show that Eq.\eqref{eq:AslamazovLarkin} provides a very good approximation if  $\epsilon \geq \epsilon_1$, see 
Eqs. (\ref{eq:rho2D},\ref{eq:rho3D}) below. Then we analyze corrections that appear at smaller values of $\epsilon$.

\section{Local noise effect}
\label{sec:LocalNoise}
Non-Gaussian effects due to interaction between fluctuating collective modes are generally known to become important 
for thermodynamics quantities in the close proximity of the critical point at $\epsilon \leq  \mathrm {Gi}$,
where Ginzburg number $ \mathrm {Gi} \sim Z^{-2/(4-d)} $.  However, it was noticed in Ref.~\cite{LO2001}, that for dynamics
quantities (in particular, for paraconductivity)  interaction corrections may become large in a parametrically broader 
range of reduced temperature $\epsilon$. In the present Section we show that similar phenomenon comes about in our model as well.
Namely, we find a rather special type of interaction corrections that affects the dependence of the relaxation
time $\tau_{GL}$ on $\epsilon$, which become relevant already at $\epsilon \leq \epsilon_1 \sim \mathrm {Gi}^{1/d}$,
whereas all static  quantities are still well-described within Gaussian approximation.

Specific kind of interaction corrections relevant at $\epsilon \leq \epsilon_1$ can be understood as a result
of local "back-action" of the order parameter (superconducting) fluctuations upon dynamics of 
individual "pseudospins"  $S_i$. Indeed, Keldysh action \eqref{eq:KeldyshAction} describes its dynamics under the fluctuating
local  ``magnetic field'' $(\Phi^x_i(t), \Phi^y_i(t), \varepsilon_i)$. Since local correlation functions of the field
$\boldsymbol{\Phi}_i(t)$  coincide with the  propagator $L(\omega, \boldsymbol{r},\boldsymbol{r}') $ calculated at
$\boldsymbol{r} = \boldsymbol{r}' = \boldsymbol{r}_i$, the action \eqref{eq:KeldyshAction} together with 
Dyson equation \eqref{eq:DysonEquation} constitute a closed set of self-consistent equations. Solution of these
equations would involve: i)  finding dynamical correlation functions of a spin $\mathrm{S}_i$ under the action of dynamic
'magnetic field' with a given correlation function of the local "noise function" 
$C_i(t-t') = \langle \Phi^\alpha_i(t)\Phi^\beta_i(t')\rangle$ ; 
ii) calculation of the propagator $L(\omega, \boldsymbol{r},\boldsymbol{r}') $ via Dyson equation; iii) self-consistent
determination of the local noise function for each site $i$. 
In general, the above scheme contains a macroscopic number of variables and thus it is untackleable.  
The problem can be grossly simplified if  site-$i$ dependent  noise function can be approximated
by a single universal function: $C_i(t-t') \to C(t-t')$.  Below in Sec.\ref{sec:RandomFieldCorrections} we will see that such an approximation is
indeed valid in the range $\mathrm{Gi} \ll \epsilon \leq \epsilon_1$. 
Currently we take it for granted and study the effect of such a transverse noise upon local spin-spin dynamics,
order parameter dynamics and, eventually, upon paraconductivity.

The key characteristic of the noise is provided by the propagator at the  coinciding points $L(\omega, \boldsymbol{r}=\boldsymbol{r}^\prime)$. This quantity itself is ultraviolet divergent (with momentum integration should be cut off at $\Lambda \sim R^{-1}$), but relevant $\omega$ and $\epsilon$-dependent part can be separated and is determined by the infrared behavior:
\begin{equation}
\label{eq:OrderParameterLocalPropagator}
L_{R}(\omega)\approx\frac{1}{8\pi\nu_{0}\xi_{0}^{d}}\begin{cases}
\ln(\Lambda^{2}\xi_{0}^{2})-\ln(\epsilon-i\omega\tau), & (2D)\\
\Lambda\xi_{0}-\sqrt{\epsilon-i\omega\tau}, & (3D)
\end{cases}.
\end{equation}
The ``local noise'' described by this propagator is small provided $\xi_0$ is large enough.

\subsection{Spin relaxation and renormalization}

The effect of the ``noise'' caused by the order parameter fluctuations  on the spin correlation function can be studied perturbatively using the Keldysh action \eqref{eq:KeldyshAction}. There are, in general, two types of such a corrections:
to a semionic propagator and to a vertex part, and we start from the first one.

\begin{figure}
	\centering
	\includegraphics[width=0.8\columnwidth]{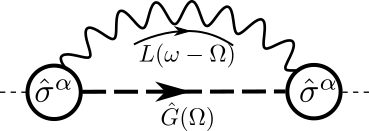}
	\caption{Semionic self-energy correction $\hat{\Sigma}(\omega)$ due to its interaction with the order parameter, which describes the ``spin noise'' effect.}
	\label{fig:SemionSelfEnergy}	
\end{figure}

The simplest diagram for the semionic propagator correction is shown on Fig. \ref{fig:SemionSelfEnergy}. 
Below we will focus only on the ``$\downarrow$'' semionic branch, as the expressions for ``$\uparrow$'' can be obtained simply by replacing $\varepsilon \to -\varepsilon$. 
The corresponding analytic expression for the retarded component of the self-energy reads:
\begin{equation}
\label{eq:SemionSelfEnergy}
\Sigma_{R}^{\downarrow}(\omega,\varepsilon)=-\frac{i}{2}\int\frac{d\Omega}{2\pi}(G_{R}^{\uparrow}(\Omega)L_{K}(\omega-\Omega)+G_{K}^{\uparrow}(\omega-\Omega)L_{R}(\Omega))
\end{equation}
For $\omega \ll T$ one can neglect the second term proportional to semionic ``distribution function'' $\mathfrak{F}(\omega)$, because the bosonic one is singular $\mathfrak{B}(\omega) \approx 2 T / \omega$. 
Under this assumption the self-energy part depends only on the simple combination of $\omega$ and $\varepsilon$, namely $\Sigma_R^{\downarrow,\uparrow}(\omega,\varepsilon) \equiv \Sigma_R(\Omega = \omega \pm \varepsilon)$ with
\begin{equation}
\label{eq:SemionicSelfEnergy}
\Sigma_{R}(\Omega)=\frac{T}{8\pi\nu_{0}\xi_{0}^{d}\Omega}\begin{cases}
\ln\frac{\epsilon-i\Omega\tau}{\epsilon}, & (2D)\\
\sqrt{\epsilon-i\Omega\tau}-\sqrt{\epsilon}, & (3D)
\end{cases}
\end{equation}

Although semions do not correspond to real quasiparticles in the system, their properties nevertheless describe the physical spin correlation function. Namely, $\Im \Sigma_R$ corresponds to the real processes of spin relaxation, and $\Re \Sigma_R$ describes renormalization of the spectrum. In the lowest order of perturbation theory, these two effects can be studied separately, and we start with the spin relaxation processes.

The spin flip rate $\gamma$, which enters the semionic Green function exactly as infinitesimal $\gamma$ did in Eq. \eqref{eq:SemionBarePropagator}, is defined by the imaginary part of $\Sigma_R$ taken on the ``mass shell'' $\omega = \varepsilon \Rightarrow \Omega = 2 \varepsilon$:
\begin{equation}
\label{eq:SemionicDecay}
\gamma(\varepsilon)\approx\frac{T}{8\pi\nu_{0}\xi_{0}^{d}}\frac{1}{\varepsilon}\cdot\begin{cases}
\arctan\frac{2\varepsilon\tau}{\epsilon}, & (2D)\\
\Im \sqrt{\epsilon+2i\varepsilon\tau} & (3D)
\end{cases}
\end{equation} 

This rate was obtained on the perturbative level, and is valid only provided the rate is small compared to the spin coherent precession frequency, $\gamma(\varepsilon) \ll \varepsilon$. This criterion clearly cannot be satisfied for all $\varepsilon$ as $\gamma(\varepsilon \to 0)$ approaches constant value. 
A new energy scale $\omega_c$ emerges, that separates spins with mainly dissipative dynamics ($\varepsilon \ll \omega_c$) from spins with coherent dynamics ($\varepsilon \gg \omega_c$). This effect can affect  paraconductivity if the energy scale
$\omega_c$  is large enough, namely, if $\omega_c \gg \epsilon T$ (note that $\omega_c$ itself can, in principle, depend on $\epsilon$). The above  criterion can be reformulated as a criterion for proximity to the transition $\epsilon \ll \epsilon_1$ with:
\begin{align}
\label{eq:rho2D}
\epsilon_{1}=&\rho^{1/2}, 
&\rho &=\frac{1}{16 \nu_0 \xi_{0}^{2}T} = \frac{g W}{8 R_0^2 T_c}, &(2D)\\
\label{eq:rho3D}
\epsilon_{1}=&\rho^{2/3}, 
&\rho &=\frac{1}{16 \sqrt{\pi} \nu_0 \xi_{0}^{3}T} = \frac{g^{3/2} W}{8 \sqrt{\pi} R_0^3 T_c}, &(3D)
\end{align}
The form of  the expression for $\omega_c$  depends on the reduced temperature $\epsilon$:
\begin{align}
\label{eq:omegac2D}
\epsilon\gg&\epsilon_{1}:& \omega_{c} = &\begin{cases}
T \rho / \epsilon & (2D)\\
T \rho \sqrt{\pi / 4\epsilon} & (3D)
\end{cases}\\
\label{eq:omegac3D}
\epsilon\ll&\epsilon_{1}:& \omega_{c} = &\begin{cases}
T\rho^{1/2}, & (2D)\\
T\rho^{2/3}, & (3D)
\end{cases}
\end{align}

For the whole analysis to be consistent, we need the
condition $\rho \ll 1 $ to be fulfilled. 
Parameter $\rho$ is inversely proportional to the coordination number, $\rho \sim 1/Z$; 
note however extra numerical factor $\sim 0.1$ in the definition of $\rho$,
which makes non-Gaussian effects smaller than one could expect.

The real part of the self-energy $\Re \Sigma_R$ renormalizes the spectral weight of the spin correlation function $\Im S_R$ in a following manner:
\begin{multline}
\Im S_{R}^{(diag)}(\omega)=\frac{1}{4}\int\frac{d\Omega}{2\pi}(\Delta G^{\downarrow}(\Omega+\omega)\Delta G^{\uparrow}(\Omega) + \{\uparrow \leftrightarrow \downarrow\}) \times \\
\times(\mathfrak{F}(\Omega)-\mathfrak{F}(\Omega+\omega))
\end{multline}
Since we are studying two effects from $\Im \Sigma_R$ and $\Re \Sigma_R$ separately, it is sufficient to substitute $\Delta G^{\uparrow,\downarrow}(\omega) = -2 \pi i \delta(\omega \pm \varepsilon - \Re \Sigma_R(\omega \mp \varepsilon))$; at low  frequencies we arrive at
\begin{multline}
\Im S_{R}^{(diag)}(\omega)=\frac{\pi\omega}{2T}\Big[(1-\Re\Sigma_{R}^{\prime}(2\varepsilon))^{-1} \times\\
\times\delta(2\varepsilon-\omega+\Re\Sigma_{R}(2\varepsilon)-\Re\Sigma_{R}(\omega))+\{\varepsilon\mapsto-\varepsilon\}\Big]
\end{multline}
This spectral weight affects relaxation time of the order parameter $\tau$ via the relation
\begin{equation}
\label{eq:TauGeneralFormula}
\omega\tau=\frac{1}{2\nu_0}\left\langle \Im S_{R}^{(diag)}(\omega)\right\rangle _{\varepsilon}
\end{equation}
However, evaluation  of Eq.\eqref{eq:TauGeneralFormula} shows that the Gaussian value for the important parameter $T \tau = \pi / 4$ remains unchanged.  We conclude that on the lowest order of perturbation theory, the effect coming from $\Re \Sigma_R$ does not affect 
the order parameter dynamics (and thus paraconductivity), and it is sufficient to focus on the spin relaxation processes only.

\begin{figure}
	\centering
	\includegraphics[width=0.8\columnwidth]{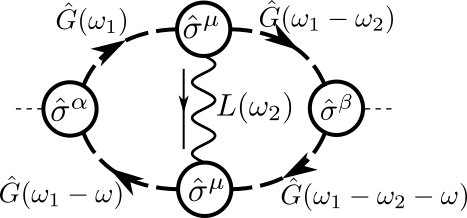}
	\caption{Vertex correction to the order parameter self-energy (Fig.\ref{fig:OrderParameterSelfEnergy}) connecting upper and lower semionic lines. This diagram vanishes due to its spin structure}
	\label{fig:OrderParameterSelfEnergyCorrection}
\end{figure}

Finally, let us focus on the vertex-part  corrections to the spin correlation function shown in 
Fig.~\ref{fig:OrderParameterSelfEnergyCorrection}; the corresponding analytic expression reads:
\begin{multline}
\delta S_{\mu \nu}^{\alpha\beta}(\omega)=-\frac{1}{4}\int\frac{d\omega_{1}}{2\pi}\frac{d\omega_{2}}{2\pi} L^{\gamma \delta}_{\lambda \rho}(\omega_{2})\Tr\Big(\check{\Gamma}_{\mu}\hat{\sigma}^{\alpha}\hat{G}(\omega_{1})\hat{\sigma}^{\gamma}\check{\Gamma}_{\lambda}\times\\
\times\hat{G}(\omega_{1}-\omega_{2})\check{\Gamma}_{\nu}\hat{\sigma}^{\beta}\hat{G}(\omega_{1}-\omega_{2}-\omega)\hat{\sigma}^{\delta}\check{\Gamma}_{\rho}\hat{G}(\omega_{1}-\omega)\Big).
\end{multline}
After working out the spin structure, one can see that two non-trivial contributions are proportional to $\Tr(\hat{\sigma}^{\alpha}\hat{\mathbb{P}}^{\uparrow}\hat{\sigma}^{\mu}\hat{\mathbb{P}}^{\downarrow}\hat{\sigma}^{\beta}\hat{\mathbb{P}}^{\uparrow}\hat{\sigma}^{\mu}\hat{\mathbb{P}}^{\downarrow})$ (and the same with $\uparrow \leftrightarrow \downarrow$); and after summation over $\mu = x,y$, these contributions exactly vanish. We conclude that lowest order non-trivial correction comes only from dressing semionic Green function in the loop, as it is  shown in Fig. \ref{fig:SemionSelfEnergy}.

\subsection{Correction to the order parameter propagator}
\label{sec:OrderParameterCorrection}

The semionic renormalization discussed above affects the spin correlation function, which enters the Dyson equation for the order parameter.
The prime effect is upon the dissipative part of the order parameter propagator $L(\omega,\boldsymbol{q})$, which is
determined  by the spectral weight of the spin correlation function, see Eq.\eqref{eq:TauGeneralFormula}. 
The major contribution to the above average over local energies $\varepsilon$ comes from $\varepsilon \sim \omega \ll T$, 
 thus the factor linear in  $\omega$  comes  just from the expansion of the Fermi distribution function,
 $\mathfrak{f}(\omega) \approx \beta \omega$. 
 This allows us to write the following formula for the important dimensionless parameter $T \tau$, which now can depend on frequency $\omega$ (we remind that paraconductivity is proportional to it, and on the Gaussian level this parameter was $\pi/4$):
\begin{equation}
\label{eq:TtauGeneral}
T\tau(\omega)=-\frac{1}{4\nu_{0}}\int\frac{d\Omega}{2\pi}\left\langle\Delta G^{\downarrow}(\Omega+\omega)\Delta G^{\uparrow}(\Omega)\right\rangle _{\varepsilon}
\end{equation}

\begin{figure}
	\centering
	\includegraphics[width=0.85\columnwidth]{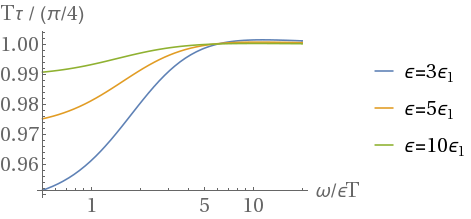}
	\includegraphics[width=0.85\columnwidth]{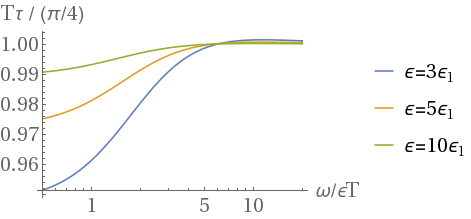}
	\caption{Frequency dependence of $T \tau(\omega)$ for 2D (upper) and 3D (lower) at $\epsilon = 3 \epsilon_1$ (blue), $\epsilon = 5 \epsilon_1$ (orange) and $\epsilon = 10 \epsilon_1$ (green) as given by Eq. \eqref{eq:TtauRelaxation}}
	\label{fig:Ttau}
\end{figure}

In the previous section we have shown that real part of semionic self-energy $\Re \Sigma_R$ does not affect the product $T\tau$, while 
 $\Im \Sigma_R$  can be accounted for by the  substitution of the propagators in the form \eqref{eq:SemionBarePropagator}
 with nonzero  $\gamma$, given by \eqref{eq:SemionicDecay}:
\begin{equation}
\label{eq:TtauRelaxation}
T\tau(\omega)=\frac{1}{4}\int\frac{d\varepsilon\cdot\gamma(\varepsilon)}{\gamma^{2}(\varepsilon)+(\varepsilon-\omega/2)^{2}}
\end{equation}
Integration can be performed numerically, plots are shown on Fig. \ref{fig:Ttau}. Striking feature of all the curves is that they exhibit non-monotonous behavior. This analysis is consistent  provided $\omega \geq \omega_c$, where the deviation of $T \tau$ from $\pi / 4$ is small.

At low frequencies $\omega \lesssim \omega_c$  kinetic term in the propagator $L(\omega,\boldsymbol{q})$ is governed by the contribution
coming from the spins with local fields $\varepsilon \lesssim \omega_c$, 
 which obey nearly dissipative dynamics,  which is difficult for analytical study.
However, it is still possible to bring up some qualitative arguments that show that the product $T\tau$ retains the same
order of magnitude and can be modified by some numerical factor $\sim 1$  only.  First we note that at $\varepsilon \ll \omega_c$
coherent contribution to spin dynamics is negligible (formally, we put here $\varepsilon =0$) and
 the only energy scale that governs  dynamics of these spins is  given by $\omega_c$.
Symmetrized spin-spin correlation  $C(t-t') = \langle \{S^+(t),S^-(t')\}\rangle = \varphi(\omega_c |t-t'|)$
where function $\varphi (z)$ decays fast at $z \gg 1$, while $\varphi(0)=1$.
 After transformation to the frequency domain, we find that  Keldysh component 
of the spin correlation function is 
$S_K^{(diag)}(\omega) \simeq \frac{1}{\omega_c} \tilde\varphi(\frac{\omega}{\omega_c})$, where
$\tilde\varphi(z)$ is some even function. Using now fluctuation-dissipation relation, one finds  that
$\Im S_{R}^{(diag)}(\omega)\simeq-\frac{\omega}{\omega_{c}T}\tilde\varphi(\frac{\omega}{\omega_{c}})$.
This form can be used now together with Eq.\eqref{eq:TauGeneralFormula}, in order to estimate the product $T\tau$.
Fraction of spins with small local energies $\varepsilon_i \leq \omega_c$ is of the order of $\sim \omega_c/W$.
Multiplying it with $\Im S_{R}^{(diag)}(\omega)$ and using Eq.\eqref{eq:TauGeneralFormula}, we find
that $T\tau \sim \tilde\varphi(0) \sim 1$.
 This qualitative argument shows that the $T \tau(\omega \ll \omega_c)$ is still a constant of the order of unity, which, however, may differ from the $\pi/4$.

\subsection{Effect on the paraconductivity}
Let us now study the  implication of the non-Gaussian effect discussed  above upon the paraconductivity. We need to consider 
the corrections to the $Q$-kernel given by Fig. \ref{fig:Qkernel}. In the leading order one should consider the same diagram, 
but with dressed order parameter Green functions, which we have studied in the previous Section. 

As it can be seen from the calculation in Sec. \ref{sec:MeanFieldConductivity}, the main contribution to the paraconductivity comes from the order parameter fluctuations with energies $\omega \tau \sim \epsilon$ and momenta $p \xi_0 \sim \sqrt{\epsilon}$.  
In the previous Section we have shown that the back-effect coming from the dynamics of  "noisy spins"  changes the 
constant $T \tau(\omega)$ at frequencies $\omega \lesssim \omega_c$ only. Thus  we conclude, that this renormalization is negligible provided $\omega_c \ll \epsilon T = |T - T_c|$, which, in turn, leads to the applicability criterion for
 Eq. \eqref{eq:AslamazovLarkin} in a form $\epsilon \geq \epsilon_1$  with $\epsilon_1$ given by Eqs.(\ref{eq:rho2D},\ref{eq:rho3D}).
 
At smaller $\epsilon$, the contribution of spins those dynamics is strongly affected by the noise, becomes dominant.
However, as we saw in the previous Subsection, this effect can hardly change the kinetic coefficient $\tau$  more substantially
than by some factor of order unity; therefore we expect Aslamazov-Larkin-type  paraconductivity,  
Eq.(\ref{eq:AslamazovLarkin}), to be  valid qualitatively even at smaller $\epsilon$, down to $\epsilon \geq \rho$.
Another types of corrections that come into play at still lower $\epsilon$, will be considered in the next Section.

\section{Other types of fluctuational corrections}
\label{sec:RandomFieldCorrections}

In the previous Section a special kind of a fluctuational correction  was demonstrated, that becomes relevant for
kinetic properties of our system in a relatively broad of reduced temperatures $\epsilon \leq \epsilon_1$,
where $\epsilon_1 \sim \rho^{1/2}$ in 2D, and  $\epsilon_1 \sim \rho^{2/3}$ in 3D.
On the other hand,  standard Ginzburg criterion for the width of fluctuation-dominated region near second-order
phase transition reads as $\epsilon \leq \epsilon_2 \equiv \mathrm{Gi} \sim Z^{-\frac{2}{4-d}} $, where $Z \sim 1/\rho$ 
is an effective number of "interacting neighbours", see Eqs.(\ref{eq:rho2D},\ref{eq:rho3D}).
Thus  we conclude that  $ \epsilon_2 \sim \epsilon_1^d \ll \epsilon_1$ for $d=2,3$.

Below in this Section we will consider some additional corrections to the Gaussian approximation of 
Sec.\ref{sec:Gaussian}, which are specific to the presence of strong disorder 
in  our model; we will show that these effects also become relevant at  $\epsilon \leq \epsilon_2 $ only.
Namely, we concentrate on the corrections to the approximation
 $\check{\Pi} = \left\langle \check{S} \right\rangle_{\varepsilon}$
 for the self-energy of the order parameter propagator $L(\omega,\boldsymbol{q})$, as defined by the 
Dyson equation \eqref{eq:DysonDisorder}.

In the calculation shown in the section \ref{sec:OrderParameterMeanField} we have studied the order parameter propagator averaged over the disorder by means of the Dyson equation \eqref{eq:DysonDisorder}, where in the leading approximation we used  the self-energy 
$\check{\Pi} = \left\langle \check{S} \right\rangle_{\varepsilon}$. The  same approximation was employed in the calculation of 
all other quantities we have studied --- including paraconductivity itself. In this Section we will study the deviations from
the results of this approximation, using  the semion  diagram technique.

\subsection{Corrections to  $L(\omega,\boldsymbol{q})$}

\begin{figure}
	\centering
	\includegraphics[width=0.8\columnwidth]{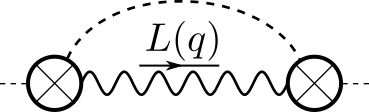}
	\caption{Correction to the averaged over disorder order parameter self-energy $\check{\Pi} - \left\langle\check{S}\right\rangle_{\varepsilon}$. Dashed ``impurity'' line corresponds to simultaneous averaging of two spin correlation functions $\check{S}$ over $\varepsilon$}
	\label{fig:OrderParameterRandomCorrection}
\end{figure}

Locator expansion for the propagator $L(\omega,\boldsymbol{q})$ averaged over distribution of $\{\varepsilon_i\}$ contains
terms of the form $\check{L}^{(0)}\check{S}\check{L}^{(0)}\check{S}\dots\check{S}\check{L}^{(0)}$. 
Previously we proceed with separate averaging of each $ \check{S} $ term in this expansion.  The first correction
to this approximation contains simultaneous averaging of two locators $ \check{S} $, as shown in 
Fig.\ref{fig:OrderParameterRandomCorrection}; the corresponding analytical expression is 
$\left\langle S^{\alpha \mu}_{R}(\omega_1) S_{R}^{\nu \beta}(\omega_2)\right\rangle_\varepsilon$.  
We present calculation of such an object in the Appendix \ref{sec:ImpurityTechnique}, making use of 
Eqs.(\ref{eq:SRstructure},\ref{eq:SRdiag},\ref{eq:SRoffdiag}).  For our  purpose it is sufficient
to consider here the limit of $\omega_{1,2} \to 0$, to obtain
\begin{equation}
\left\langle (S_{R}^{(diag)}(0))^2\right\rangle_\varepsilon \approx \frac{14 \zeta(3)}{\pi^2 WT}
\label{2S}
\end{equation}
The structure of the correction shown in Fig. \ref{fig:OrderParameterRandomCorrection} appears to be diagonal
in the $(\alpha,\beta)$ space,
$\delta \Pi_R^{\alpha \beta}(\omega) = (\delta^{\alpha \beta} - n^\alpha n^\beta) \delta \Pi_R(\omega)$,
 and the whole correction to the self-energy is given by
\begin{equation}
\delta\Pi_{R}(\omega) = 
L_{R}(\omega)\left\langle (S_{R}^{(diag)}(0))^{2}\right\rangle
\label{deltaPi}
\end{equation}

The static ($\omega=0$) contribution to $\delta\Pi_R(\omega)$  corresponds to the renormalization of $T_c$, which was already studied in the Appendix \ref{sec:AppendixSCBA} (see Eq. \ref{eq:TcRenormalization} and comments below).  The frequency-dependent part at  $\omega \tau \ll \epsilon$ contains a singularity
at small $\epsilon$:

\begin{equation}
\label{dP}
\delta\Pi_R(\omega) - \delta\Pi_R(0) \approx -\frac{i\omega\tau}{4\pi T\xi_{0}^{d}}\times\begin{cases}
\frac{14\zeta(3)}{\pi^{2}\epsilon}, & (2D)\\
\frac{7\zeta(3)}{\pi^{2}\sqrt{\epsilon}}, & (3D)
\end{cases}
\end{equation}
which should be compared with the bare $\omega$-dependent term in $L^{-1}(\omega,\boldsymbol{q})$, see 
Eq.\eqref{eq:OrderParameterPropagator}.
Then we find that the correction  \eqref{dP} is small provided
\begin{equation}
\label{eq:FluctuationsApplicability}
\begin{cases}
\epsilon\gg\frac{W}{T\xi_{0}^{2}}\sim\rho, & (2D)\\
\epsilon\gg\left(\frac{W}{T\xi_{0}^{3}}\right)^{2}\sim\rho^{2}, & (3D)
\end{cases}
\end{equation} 
which coincides with the usual Ginzburg criterion discussed in the beginning of this Section.

\subsection{Spatial fluctuations of the conductivity}
\label{sec:ConductivityFluctuations}

 It was assumed implicitly  during the calculation of paraconductivity in Sec.\ref{sec:MeanFieldConductivity}  (and further discussion in Sec. IV C)
 that  conductivity is uniform through the system and thus can be characterized
as the kernel $\sigma(\boldsymbol{r}-\boldsymbol{r'})$ in the linear relation between current density and electric field,
$\boldsymbol{j}(\boldsymbol{r}) = \int  \sigma(\boldsymbol{r}-\boldsymbol{r'}) \boldsymbol{E}(\boldsymbol{r'}) d^3r' $. 
In the disordered medium, conductivity   contains spatial fluctuations, so that the kernel becomes a function of two
coordinates separately, $\sigma(\boldsymbol{r}-\boldsymbol{r'}) \to \sigma(\boldsymbol{r}, \boldsymbol{r'})$.
In order to satisfy current conservation law, $\partial_{\alpha} j_\alpha = 0$, 
with the current given by $j_{\alpha}(\boldsymbol{r})=
\int\sigma(\boldsymbol{r},\boldsymbol{r}^{\prime})E_{\alpha}(\boldsymbol{r}^{\prime})d\boldsymbol{r}^{\prime}$,
the local electric field $E_\alpha$ must fluctuate in space:
\begin{equation}
\delta E_{\alpha}(\boldsymbol{r})=-\frac{1}{d\overline{\sigma}}\overline{E}_{\alpha}\int\delta
\sigma(\boldsymbol{r},\boldsymbol{r}^{\prime})d\boldsymbol{r}^{\prime},
\end{equation}
It results~\cite{LandauLifshits} in the additional
contribution to the average conductivity of the form
$$\delta\sigma = -\frac{1}{d} K(\textbf{r} = 0)$$
where correlation function $K(\boldsymbol{r}- \boldsymbol{r'})$ is defined as follows:
\begin{equation}
\label{Kdef}
K(\boldsymbol{r}-\boldsymbol{r}^{\prime})=\frac{1}{\bar{\sigma}^{2}}\int d\boldsymbol{x}d\boldsymbol{y}\left\langle \delta\sigma(\boldsymbol{r},\boldsymbol{x})\delta\sigma(\boldsymbol{r}^\prime,\boldsymbol{y})\right\rangle 
\end{equation}
\begin{figure}
	\centering
	\includegraphics[width=0.8\columnwidth]{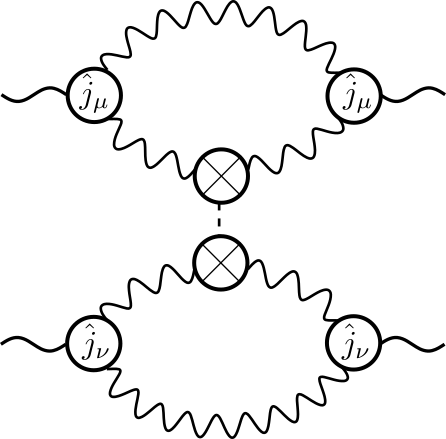}
	\caption{Lowest order diagram representing the fluctuations of the conductivity $K(\boldsymbol{r}-\boldsymbol{r}^\prime)$}
	\label{fig:ConductivityFluctuations}
\end{figure}
Below we will calculate this correlation function $K(\boldsymbol{r}- \boldsymbol{r'})$.  The diagram of  the lowest order is shown in Fig. \ref{fig:ConductivityFluctuations}. This  diagram consists of two parts: two independent loop integrals, which are similar to the $Q$-kernel given by Eq. \eqref{eq:QkernelLoop} and which we denote as ${\cal R}^{i j}(\omega,\boldsymbol{q})$, and an ``impurity line'', which to the leading order can be taken in the static limit $\left\langle (S_{R}^{(diag)}(0))^{2}\right\rangle$. Since one can put an ``impurity'' either on upper or lower Green function, which corresponds to replacement $q \mapsto -q$ in the expression for ${\cal R}^{\mu \nu}(\omega,\boldsymbol{q})$, there are, in total, four terms in the expression for the
 conductivity fluctuations:
\begin{equation}
K(\boldsymbol{q})=\left\langle (S_{R}^{(diag)}(0))^{2}\right\rangle \left[\frac{i\partial_{\omega}({\cal R}^{ii}(\omega,\boldsymbol{q})+{\cal R}^{ii}(\omega,-\boldsymbol{q})}{d\sigma_{0}}\right]_{\omega=0}^{2}
\end{equation}
The explicit calculation of the ${\cal R}$ is provided in the Appendix \ref{sec:AppendixFluctuations}; using the dimensionless function ${\cal F}(\boldsymbol{Q} = \boldsymbol{q} \xi_0 / \sqrt{\epsilon})$ and substituting bare value of conductivity given by Eq. \eqref{eq:ALArbitraryD}, we arrive at the following general expression:
\begin{equation}
K(\boldsymbol{q})=\frac{56\zeta(3)W}{\pi^{2}\sigma_{d}^{2}T\epsilon^{2}}{\cal F}^{2}(\boldsymbol{Q}).
\end{equation}
The relative scale of spatial fluctuations of the conductivity is thus given by $K(\textbf{r} = 0)$.
 Both asymptotics \eqref{eq:AsymptoticF2D} and \eqref{eq:AsymptoticF3D} show that the integral that defines
$K(\textbf{r} = 0)$ is convergent, finally it yields:
\begin{equation}
K(\boldsymbol{r}=0)=c \frac{W}{T\epsilon^{2-d/2}\xi_{0}^{d}},
\label{Kend}
\end{equation}
with the  prefactor $c$, which can be obtained  numerically:
\begin{equation}
c=\frac{56\zeta(3)}{\pi^{2}\sigma_{d}^{2}}\int\frac{d^{d}\boldsymbol{Q}}{(2\pi)^{d}}{\cal F}^{2}(\boldsymbol{Q})=\begin{cases}
0.283, & (2D)\\
0.099, & (3D)
\end{cases}
\end{equation}

Now let us discuss the obtained result. The Aslamazov-Larkin formula \eqref{eq:AslamazovLarkin} works only provided the correction $K(\textbf{r} = 0) \ll 1$. The result is essentially the same as the one obtained in the previous Section: 
the correction is small provided Eq. \eqref{eq:FluctuationsApplicability} holds.

\section{Conclusions}
\label{sec:Conclusions}

We have shown in this paper that  fluctuational conductivity effect, originally predicted by L.Aslamazov and A.Larkin 50 years 
ago, remains nearly the same in the case of strongly pseudo-gaped superconductors with just absent single-electron density 
of states. The role of single-electron states is taken over by  the localized electron pairs, and the effect of that replacement
reduces just to the factor of 2 change of the numerical coefficient $\sigma_d$  in Eqs.
(\ref{eq:ALArbitraryD},\ref{eq:AslamazovLarkin})  w.r.t. to the classical  Aslamazov-Larkin result, 
while power-law dependence of paraconductivity on $\epsilon=\ln(T/T_c)$ remains the same.
Our results were derived under the assumption that hopping of (initially) localized pairs occurs with a
large effective "coordination number"  $Z  \sim \rho^{-1}$, see Eq.(\ref{eq:rho2D},\ref{eq:rho3D}).

Universal character of the AL  paraconductivity (especially, in 2D) makes it  convenient experimental tool for
determination of the critical temperature when  $R(T)$ dependence is of considerable width, like it occurs in
strongly disordered superconductors.  For this reason, the issue of universality of the value of numerical coefficient $\sigma_d$ 
is of interest.  First, we note that it does not depend upon the shape of the local energy distribution function
$P(\varepsilon)$ as long as it is flat on the scale of  very small $\varepsilon \sim T_c$.  
Some nontrivial structure at this energy scale in the effective distribution $P(\varepsilon)$ may come about 
in the generalized model where long-range interaction of the
type of $S_i^z U(\boldsymbol{r}_i - \boldsymbol{r}_j) S_j^z$ is included, that can be traced back to the Coulomb interaction
between charges of localized pairs.  The effect of such an interaction will be studied separately.

Since our condition of a very large pseudogap $\Delta_P \gg T_c$ may be found too restrictive in applications, one
might be interested in generalization of our result for moderate value of $\Delta_P \sim T_c$. That can be done in
a rather straightforward way, once we note that the whole issue of the coefficient $\sigma_d$ in Eq.(\ref{eq:AslamazovLarkin})
is controlled by the expansion of the effective spin distribution function $\mathfrak{f}(\omega) = \tanh \beta \omega$ over
small $\omega$. In the standard TDGL theory \cite{VarlamovLarkin} for disordered superconductors, the  fermionic
distribution function $ f(\omega) = \tanh \frac{\beta \omega}{2}$ stays instead of $\mathfrak{f}(\omega) $, thus making the
coefficient in front of $T\tau$ twice smaller than in our problem, see Eq.(\ref{GLparameters}).
For the general case of $\Delta_P \sim T_c$ we can use an observation presented in the Appendix B to the paper~\cite{t2}: 
for an arbitrary $\Delta_P/T$, a generalized distribution function is
\begin{equation}
\mathfrak{f}(\omega,\Delta_P) = \frac{\sinh \beta\omega}{\cosh\beta\omega + e^{-\Delta_P/T}}
\nonumber
\end{equation}
which interpolate between $\tanh\frac{\beta \omega}{2}$ and  $\tanh \beta \omega$ upon increase of $\Delta_P/T$.
As a result,  for a generic $\Delta_P$ values, the enhancement factor in $\sigma_d$,
 w.r.t. to the standard AL result, is given by
$ 2/(1 + e^{-\Delta_P/T_c}) $, i.e. it quickly becomes close to 2 for moderate $\Delta_P/T \geq 1.5$.

All the above discussion refers to the Gaussian fluctuation region, $\epsilon \geq \epsilon_1$, see 
Eqs.(\ref{eq:rho2D}),(\ref{eq:rho3D}).  At smaller $\epsilon$ non-linear corrections to the dynamics of the order parameter
becomes important, they are discussed in Sec. IV.  However, we present  the arguments  that power-law character
of $\sigma_{AL}(\epsilon)$ dependence is not changed due to these "local noise" effects, while the coefficient
$\sigma_d$ becomes somewhat different.  Even more close to $T_c$, at $\epsilon \leq\epsilon_2 = \mathrm{Gi}$, all types
of fluctuational corrections becomes relevant,  which makes calculation of $\sigma_{AL}(\epsilon)$ difficult.
Moreover, in this close proximity of $T_c$, conductivity becomes spatially inhomogenuous, 
as evidenced by Eq.\eqref{Kend}.

Specific feature of fluctuational conductivity in superconductors close to SIT is that it may  much exceed bare 
(unrelated to superconducting correlations) conductivity already in the region of $\epsilon \geq \epsilon_1$ where
Gaussian approximation is valid.  This is due to the absence in our case of the standard Drude contribution of the 
normal-metal type.  Instead, Aslamazov-Larkin paraconductivity competes with hopping conductivity of individual
electrons, that is further suppressed at $T \ll \Delta_P$.

We are grateful to  Lev Ioffe  for useful discussions. This research was partially supported by the 
Russian Science Foundation grant \# 14-42-00044, by the Basic research program of the HSE,  
and  by the grant from the Basis Foundation (I.P.).

\appendix
\section{Mean field approximation and finite $R$ effects on $T_c$}
\label{sec:AppendixSCBA}
This Appendix is devoted to analytical and numerical study of the critical temperature $T_c$ at the mean-field level. In  Section \ref{sec:MeanFieldStatic} we have formulated the following condition for the appearance of order parameter: the largest eigenvalue of the matrix $J_{ij} \sqrt{\eta_i \eta_j}$ should be larger than unity. This criterion was then solved in the limit
 $R \to \infty$ yielding Eq. \eqref{eq:TcMeanField}.  Here we consider leading corrections to this result at large $R$.

We start our analysis with analytical treatment of the spectrum (DoS) of matrix $J_{ij} \sqrt{\eta(\varepsilon_i) \eta(\varepsilon_{j})}$, averaged over the distribution $P(\varepsilon)$. For this purpose we express the DoS in terms of Green function $\hat{G}_E=(E-\hat{\eta}^{1/2}\hat{J}\hat{\eta}^{1/2}+i0)^{-1}$ as $\nu(E) = -\frac{1}{\pi N}\Tr\hat{G}_E$, and expand it in Dyson series. The latter can be rewritten more conveniently in terms of auxillary matrix $\hat{F}_{E}=\hat{\eta}^{-1/2}\hat{G}_{E}\hat{\eta}^{1/2}\hat{J}$:
\begin{equation}
\hat{F}_{E}=\hat{F}_{E}^{(0)}+\hat{F}_{E}^{(0)}\hat{\eta}\hat{F}_{E}^{(0)}+\hat{F}_{E}^{(0)}\hat{\eta}\hat{F}_{E}^{(0)}\hat{\eta}\hat{F}_{E}^{(0)}+\dots,
\end{equation}
with $\hat{F}_{E}^{(0)}=\hat{J}/(E+i0)$. Under the assumption of large radius $R$, we can apply an ordinary impurity diagram technique utilizing equation $\left\langle \eta_{i}\eta_{j}\right\rangle =\delta_{ij}\left\langle \eta^{2}\right\rangle +(1-\delta_{ij})\left\langle \eta\right\rangle ^{2}$. The first approximation for the self-energy corresponds to trivial mean-field analysis performed in the Section \ref{sec:MeanFieldStatic} and reads $\hat{\Sigma}^{(1)} = \left\langle\eta\right\rangle$. In order to study the DoS near the spectrum edge, we utilize the self-consistent Born approximation (SCBA) and consider following self-energy correction:
\begin{equation}
\hat{\Sigma}^{(2)}=\left\langle \left\langle \eta^{2}\right\rangle \right\rangle \cdot(\hat{F}_{E})_{ii}
\end{equation}
In the momentum representation, the Dyson equation for the SCBA then reads:
\begin{equation}
\label{eq:SCBA}
F_{E}^{-1}(\boldsymbol{q})=J(\boldsymbol{q})^{-1}(E+i0)-\left\langle \eta\right\rangle -\left\langle \left\langle \eta^{2}\right\rangle \right\rangle F(E),
\end{equation}
with $F(E)=\int(d\boldsymbol{q})F_{E}(\boldsymbol{q})$. This allows us to write a single self-consistency equation for $F(E)$: 
\begin{equation}
\label{eq:SCBAF}
F(E)=\int\frac{d^d\boldsymbol{q} / (2\pi)^d}{J^{-1}(\boldsymbol{q})(E+i0)-\left\langle \eta\right\rangle -\left\langle \left\langle \eta^{2}\right\rangle \right\rangle F(E)}
\end{equation}

The next step is to express the density of states in terms of the function $F(E)$. First we note that $\hat{G}_{E}=\hat{\eta}^{1/2}\hat{F}_{E}\hat{J}^{-1}\hat{\eta}^{-1/2}$, and thus $\Tr\hat{G}_{E}=\Tr(\hat{F}_{E}\hat{J}^{-1})$.
In the UV limit $q \to \infty$ we have $J(\boldsymbol{q}) \to 0$, which leads to the delta-peak at zero energy. Since we are studying the edge of the spectrum, we can subtract the value $(E + i0)^{-1}$ and focus at $E > 0$. Utilizing then equation for $F(E)$, we obtain following general expression for the DoS:
\begin{equation}
\nu(E>0)=-\frac{1}{\pi E}{\rm Im}\left(F(E)\left[\left\langle \eta\right\rangle +\left\langle \left\langle \eta^{2}\right\rangle \right\rangle F(E)\right]\right)
\end{equation}

We now proceed with solving the equation \eqref{eq:SCBAF}. We switch to dimensionless momentum $Q = q R$ and dimensionless variables:
\begin{equation}
\lambda=\frac{E}{J\left\langle \eta\right\rangle }-1,\quad \Psi(\lambda) = F(E) \left\langle \eta\right\rangle R^{d},\quad j(\boldsymbol{Q}) = J(\boldsymbol{q})/J
\end{equation}
leaving us with the single small dimensionless parameter, which controls the SCBA:
\begin{equation}
\alpha=\frac{\left\langle\left\langle \eta^{2}\right\rangle\right\rangle }{R^{d}\left\langle \eta\right\rangle ^{2}}=\frac{1}{R^{d}}\left[\frac{14\zeta(3)}{\pi^{2}}\frac{\beta W}{\ln^{2}\frac{4e^{\gamma}\beta W}{\pi}}-1\right]\sim\frac{g^{2}e^{1/g}}{R^{d}} \ll 1.
\end{equation}
The dimensionless form of the equation \eqref{eq:SCBAF} then reads:
\begin{equation}
\label{eq:PsiExact}
\Psi(\lambda)=\int\frac{d^d\boldsymbol{Q}/ (2\pi)^d \cdot j(\boldsymbol{Q})}{\lambda+i0+1-j(\boldsymbol{Q})-\alpha j(\boldsymbol{Q})\Psi(\lambda)},
\end{equation}
and DoS is expressed in terms of $\Psi$-function as follows:
\begin{equation}
\label{eq:DOSSCBA}
\nu(E)=-\frac{1}{\pi ER^{d}}{\rm Im}\left[\Psi(\lambda)+\alpha\Psi^{2}(\lambda)\right].
\end{equation}

The long-wavelength limit $j(\boldsymbol{Q}) = 1 - Q^2$  is sufficient for the study of the DoS behavior near the spectrum edge
$E \approx J \left\langle \eta\right\rangle$, that is $|\lambda| \ll 1$. Performing momentum expansion and focusing for the sake of simplicity on 2D case, where the integral is logarithmic, we arrive at following equation:
\begin{equation}
\Psi(\lambda)\approx\frac{1}{4\pi}\ln\frac{c}{\lambda + i0-\alpha\Psi(\lambda)},
\end{equation}
where $c$ is a constant of order of unity depending on the UV behavior of $j(\boldsymbol{Q})$. In the limit $R \to \infty$, that is $\alpha = 0$, this equation leads to the step-like DoS with the sharp edge at $\lambda = 0$, 
$\nu(E)\approx\theta(J\left\langle \eta\right\rangle -E) / 4\pi R^2$. Finite but small $\alpha$ rounds out the step 
leading to the square-root singularity at the slightly shifted edge:
\begin{equation}
\label{eq:DOSSCBAedge}
\nu(E)\approx\frac{1}{\pi ER^{d}}\sqrt{\frac{\lambda_{G}-\lambda}{2\pi\alpha}},
\end{equation}
with 
\begin{equation}
\label{eq:TcRenormalization}
\lambda_{G}=\frac{\alpha}{4\pi}\ln\frac{4\pi ec}{\alpha}.
\end{equation}
The shift of the spectrum edge leads to the renormalization of the coupling constant $J_\textrm{eff} = J(1+\lambda_G)$ in expression for $T_c$, Eq.\eqref{eq:TcMeanField}, thus  increasing the critical temperature slightly.

To support this calculation, we have performed numerical analysis of the spectrum of corresponding random matrix. The temperature $T = \beta^{-1}$ was taken close to the mean-field value of the critical temperature \eqref{eq:TcMeanField}, so that spectrum edge is estimated to be close to unity. The $J_{ij}$ matrix was taken Gaussian, so that its Fourier transform has the form $J(\boldsymbol{q}) = J \exp(-q^2 R^2)$; in that case the integration in Eq. \eqref{eq:PsiExact} can be performed explicitly leaving the single algebraic equation for $\Psi$, which is then solved numerically to obtain the analytic fitting curve. The amount of disorder realization varied from $\sim$30000 (for smallest system) to $\sim$6000 for largest one.

\begin{figure}
	\includegraphics[width=0.8\columnwidth]{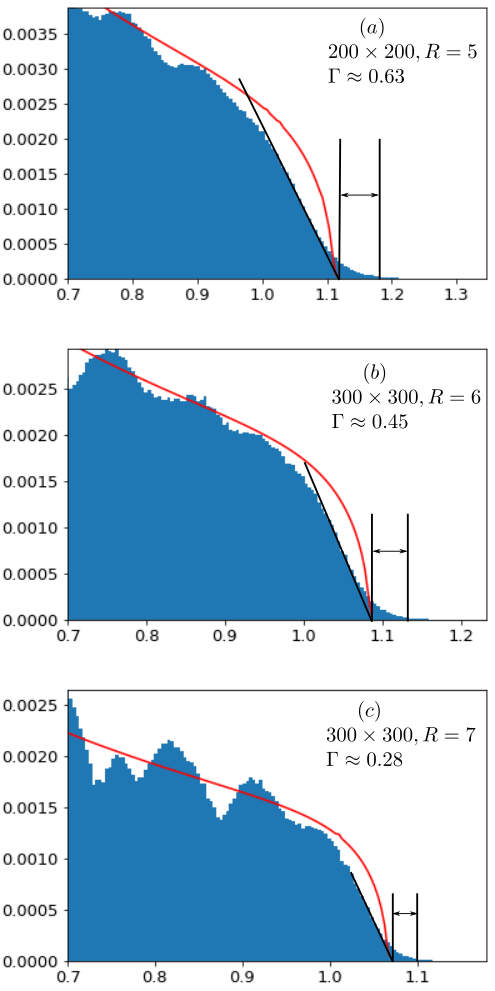}
	\caption{DoS $\nu(E)$ for 2D system with parameters $W = 3$, $J = 1$, which corresponds by Eq. \eqref{eq:TcMeanField}, $T_c^{-1} \approx 60$. Red curve: solution of the SCBA equation, \eqref{eq:PsiExact}, and substituting the solution to \eqref{eq:DOSSCBA}. The}
	\label{fig:DOS}
\end{figure}

\begin{figure}
	\includegraphics[width=0.8\columnwidth]{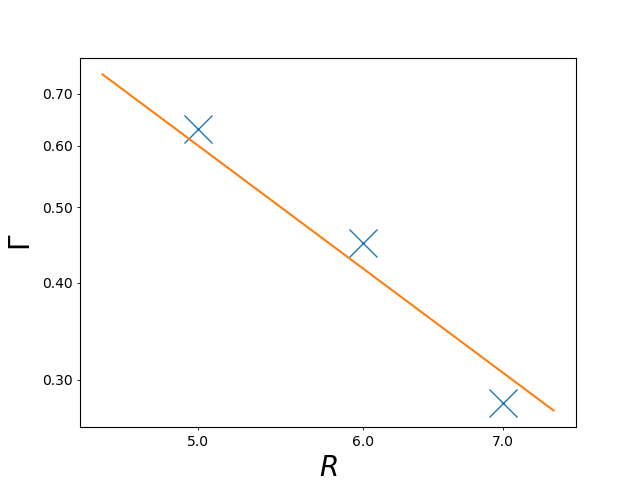}
	\caption{$R$-dependency of the width of the tail extracted from Fig. \ref{fig:DOS}. Line corresponds to $R^{-2}$}
	\label{fig:tails}
\end{figure}

The typical DoS pictures shown on Fig. \ref{fig:DOS} consists of the ``main body'' of the DoS, which is fitted by the SCBA formula reasonably well, and the exponential tail of localized states which always arise when one deals with random matrices. The oscillatory behavior is due to finite-size effects and momentum quantization in a system of finite size; they tend to increase upon increasing $R$ and decreasing $L$. The superconductivity  appears when the mobility edge separating localized and delocalized states crosses unity eigenvalue. Clearly, the edge of the spectrum is larger than unity, and the SCBA result, Eq. \eqref{eq:DOSSCBAedge} gives the better estimation of the position of the edge as well as the whole curve.

The width of the tail $\Gamma$ is known to be related to the Ginzburg number $\mathrm{Gi} \sim \rho^{2 / (4-d)} \propto R^{-2d/(4-d)}$. To support this claim, we performed numerical simulations for system with various $R$ and estimated the $R$-dependency of the width of the tail. The beginning of the tail was determined by the intersection of the tangent line to the curve in the inflection point with the $x$-axis, see Fig. \ref{fig:DOS}. The best fit of the $\Gamma(R)$ dependency is shown on Fig. \ref{fig:tails}, and corresponds to $\Gamma(R) \sim R^{-2}$, which agrees with the prediction for $d = 2$.

\section{Keldysh diagram technique for pseudospins}
\label{sec:AppendixKeldysh}
In this Appendix we will derive the Keldysh action and rules for diagram technique used to describe the pseudospin model \eqref{eq:PseudospinHamiltonian} and its semionic representation \eqref{eq:SemionHamitlonian} following Kiselev and Opperman \cite{KiselevOppermann}. We introduce the standard Keldysh time contour $C = (-\infty,\infty)\cup(\infty,-\infty)$ for the model and introduce the following action for the semions:
\begin{multline}
i S[\bar{\psi}, \psi] = i\int_C dt \left(\bar{\psi}\hat{G}^{-1}\psi+\frac{1}{4}(\bar{\psi}\hat{\sigma}^{\alpha}\psi)\hat{J}(\bar{\psi}\hat{\sigma}^{\alpha}\psi)\right),
\end{multline}
where $\hat{G}^{-1}=i\partial_{t}+\varepsilon_{i}\sigma^{z}$ is diagonal matrix in real space, and implied summation over coordinates. We introduce the two-component real Hubbard-Stratanovich field $\Phi = (\Phi^x, \Phi^y)$ with the following action:
\begin{equation}
iS[\Phi]=-i\int_{C}dt\Phi^{\alpha}\hat{J}^{-1}\Phi^{\alpha},
\end{equation}
and perform a shift to decouple four-semion interaction $\Phi^{\alpha}\mapsto\Phi^{\alpha}-\frac{1}{2}\hat{J}\bar{\psi}\sigma^{\alpha}\psi$, arriving at following action:
\begin{equation}
iS[\bar{\psi}, \psi, \Phi] = i\int_C dt\left(-\Phi^{\alpha}\hat{J}^{-1}\Phi^{\alpha}+\bar{\psi}(\hat{G}^{-1}+\hat{\sigma}^{\alpha}\Phi^{\alpha})\psi\right).
\end{equation}
This action effectively describes spins-$1/2$ lying in a fluctuating magnetic field $(\Phi_i^x(t), \Phi_i^y(t), \varepsilon_i)$, whose dynamics itself is coupled to the spins via the interaction vertex. This is thus a clear generalization of a simple static mean-field model described in Section \ref{sec:MeanFieldStatic}.

The next step is to separate fields lying on the upper and lower parts of the Keldysh contour as $\Phi = (\Phi_{+}, \Phi_{-})$ (and similarly for $\psi$), and introduce a Keldysh rotation switching to ``classical'' and ``quantum'' bosonic fields: $\Phi^\prime = (\Phi_{cl}, \Phi_q)$; and their analog for fermions: $\psi^\prime = (\psi_1, \psi_2)^T$, $\bar{\psi}^\prime = (\bar{\psi}_1, \bar{\psi}_2)$, via the following relations:
\begin{equation}
\Phi = \check{O} \Phi^\prime,\quad \psi = \check{O} \psi^\prime,\quad \bar{\psi} = \bar{\psi}^\prime \check{O}\check{\tau}_z,
\end{equation}
with matrix $\check{O} = (\check{\tau}_x + \check{\tau}_z) / \sqrt{2}$ and $\check{\tau}_\alpha$ being Pauli matrices acting in Keldysh space. This rotation yields following Keldysh structure of the propagators:
\begin{equation}
\check{L} = i \left\langle \Phi \Phi^T \right \rangle = \begin{pmatrix}L_{K} & L_{R}\\
L_{A} & 0
\end{pmatrix},
\end{equation}
\begin{equation}
\check{G} = -i \left\langle \psi \bar{\psi}\right\rangle = \begin{pmatrix}G_{R} & G_{K}\\
0 & G_{A}
\end{pmatrix}.
\end{equation}

Finally, the ``rotated'' action is given by Eq. \eqref{eq:KeldyshAction}. In principle, one can perform Gaussian integration over semionic degrees of freedom and obtain the effective action describing only the order parameter dynamics:
\begin{equation}
i S[\Phi] = -i\int dt \Phi^{\alpha}\hat{J}^{-1}\check{\tau}_{x}\Phi^{\alpha}+\Tr\ln\left(\hat{G}^{-1}+\frac{1}{\sqrt{2}}\check{\Gamma}_{i}\hat{\sigma}^{\alpha}\Phi_{i}^{\alpha}\right).
\end{equation}

We remind that the order parameter fields $\Phi \equiv \Phi^\alpha_\mu(\boldsymbol{r}_i)$ has the following indices: ``spin space'' $\alpha \in (x,y)$, Keldysh space $\mu \in (cl, q)$ and real space $\boldsymbol{r}_i$; 
while the semionic fields $\psi \equiv \psi_{\sigma, \mu}(\boldsymbol{r}_i)$ reside in ``semionic pseudospin space'' $\sigma \in \{\uparrow, \downarrow\}$, Keldysh space $\mu \in \{1,2\}$ and real space $\boldsymbol{r}_i$.

\section{``Impurity'' diagram technique}
\label{sec:ImpurityTechnique}
The aim of this appendix is to develop an ``impurity''-like diagram technique which is used in Section \ref{sec:RandomFieldCorrections} to study the deviations from the mean-field approximation presented in Section \ref{sec:Gaussian} due to the averaging over the distribution of $\{\varepsilon_i\}$. The key element of the diagram technique that depended on the $\varepsilon$ is the ``crossed circle'' presented on Fig. \ref{fig:OrderParameterSelfEnergy}, which represents spin correlation function $S_R^{\alpha \beta}(\omega)$. Upon averaging the e.g. Dyson series \eqref{eq:DysonEquation}, the next non-trivial object arising is simultaneous averaging of two spin correlation functions corresponding to the single site $\left\langle S^{\alpha \mu}_{R}(\omega_1) S_{R}^{\nu \beta}(\omega_2)\right\rangle_\varepsilon$, which corresponds to the ``impurity line'' connecting two crossed circles in our diagram technique.

We proceed with the calculation of analytic expression for ``impurity line'' utilizing the spin structure \eqref{eq:SRstructure} and expressions \eqref{eq:SRdiag} and \eqref{eq:SRoffdiag}. The cross-term $\left\langle S_{R}^{(diag)}(\omega_{1})S_{R}^{(off)}(\omega_{2})\right \rangle_\varepsilon$  drops out due to its parity, while non-zero terms for $\omega_{1,2} \ll T$ yield:
\begin{multline}
\left\langle S_{R}^{(diag)}(\omega_{1})S_{R}^{(diag)}(\omega_{2})\right\rangle_\varepsilon =\\
=\left\langle\frac{\mathfrak{f}^{2}(\varepsilon)\varepsilon^{2}}{((\omega_{1}/2+i0)^{2}-\varepsilon^{2})((\omega_{2}/2+i0)^{2}-\varepsilon^{2})}\right\rangle_{\varepsilon} \approx \\	
\approx \frac{14 \zeta(3)}{\pi^2 WT} + \frac{i\pi}{4 WT^{2}}\frac{\omega_{1}^{2}+\omega_{2}^{2}+\omega_{1}\omega_{2}}{\omega_{1}+\omega_{2}+i0},
\end{multline}
\begin{multline}
\left\langle S_{R}^{(off)}(\omega_{1})S_{R}^{(off)}(\omega_{2})\right\rangle_\varepsilon = \\
= \frac{\omega_{1}\omega_{2}}{4}\left\langle \frac{\mathfrak{f}^{2}(\varepsilon)}{((\omega_{1}/2+i0)^{2}-\varepsilon^{2})((\omega_{2}/2+i0)^{2}-\varepsilon^{2})}\right\rangle_{\varepsilon} \approx  \\
\approx \frac{i\pi}{4WT^{2}}\frac{\omega_{1}\omega_{2}}{\omega_{1}+\omega_{2}+i0},
\end{multline}

\section{Calculation of conductivity correlation function}
\label{sec:AppendixFluctuations}
Here we briefly discuss the calculation of the conductivity fluctuations $\left\langle \delta\sigma(\boldsymbol{r},\boldsymbol{x})\delta\sigma(\boldsymbol{r}^\prime,\boldsymbol{y})\right\rangle$ in the lowest order of perturbation theory given by the diagram shown on Fig. \ref{fig:ConductivityFluctuations}. The analytic expression for the loop integrals ${\cal R}^{i j}$ appearing in the Section \ref{sec:ConductivityFluctuations}  reads as follows:
\begin{widetext}
\begin{multline}
{\cal R}^{i j}(\omega,\boldsymbol{q})=i\frac{8\xi_{0}^{4}}{W^{2}}\int\frac{d\Omega}{2\pi}\frac{d^d\boldsymbol{p}}{(2\pi)^d}p^{i}p_{+}^{j}\Bigg[\mathfrak{B}(\Omega_{-})L_{R}(\Omega_{+},\boldsymbol{p}_{+})(L_{R}(\Omega_{-},\boldsymbol{p}_{+})L_{R}(\Omega_{-},\boldsymbol{p}_{-})-L_{A}(\Omega_{-},\boldsymbol{p}_{+})L_{A}(\Omega_{-},\boldsymbol{p}_{-}))+\\
+\mathfrak{B}(\Omega_{+})(L_{R}(\Omega_{+},\boldsymbol{p}_{+})-L_{A}(\Omega_{+},\boldsymbol{p}_{+}))L_{A}(\Omega_{-},\boldsymbol{p}_{+})L_{A}(\Omega_{-},\boldsymbol{p}_{-}))\Bigg]
\end{multline}
In the low-frequency limit we take $\mathfrak{B}(\Omega) \approx 2T / \Omega$, and perform integration over energy $\Omega$ using residues, arriving at:
\begin{equation}
{\cal R}^{i j}(\omega,\boldsymbol{q})=16WT\xi_{0}^{4}\int\frac{d^{d}\boldsymbol{p}}{(2\pi)^{d}}p^{i}p_{+}^{j}\frac{\epsilon+(\boldsymbol{p}^{2}+\boldsymbol{q}^{2}/4)\xi_{0}^{2}-i\omega\tau/4}{(\epsilon+\boldsymbol{p}_{-}^{2}\xi_{0}^{2})(\epsilon+\boldsymbol{p}_{+}^{2}\xi_{0}^{2})\left(\epsilon+\boldsymbol{p}_{+}^{2}\xi_{0}^{2}-i\omega\tau/2\right)\left(\epsilon+(\boldsymbol{p}^{2}+q^{2}/4)\xi_{0}^{2}-i\omega\tau/2\right)}
\end{equation}
This integral is taken at finite external momentum and thus it can have non-trivial tensor structure. We are interested in the diagonal conductivity, which is $\delta\sigma = \frac{1}{d}\sigma^{ii}$. The next step is to make momentum integration dimensionless introducing $\boldsymbol{P}=\boldsymbol{p}\xi_{0}/\sqrt{\epsilon}$, expand it in frequency, take trace ${\cal R}^{i i} / d\equiv {\cal R}$ and introduce dimensionless function ${\cal F}(\boldsymbol{Q})$:
\begin{equation}
i\partial_{\omega}{\cal R}(\omega=0,\boldsymbol{Q}) = -\frac{W}{\xi_{0}^{d-2}\epsilon^{3-d/2}}\frac{\pi}{d}\int\frac{d^{d}\boldsymbol{P}}{(2\pi)^{d}}\frac{\boldsymbol{P}^2 + \boldsymbol{P}\boldsymbol{Q}/2}{(1+\boldsymbol{P}_{-}^{2})(1+\boldsymbol{P}_{+}^{2})^{2}}\left(\frac{1}{1+P^{2}+\boldsymbol{Q}^{2}/4}+\frac{2}{1+\boldsymbol{P}_{+}^{2}}\right) \equiv -\frac{W}{\xi_{0}^{d-2}\epsilon^{3-d/2}} {\cal F}(\boldsymbol{Q})
\end{equation} 
\end{widetext}
We now switch to the calculation of this function in arbitrary spatial dimensionality.
\paragraph{2D case}
Using substitution $a = 1 + P^2 + Q^2/4$, we perform angular averaging arriving at:
\begin{equation}
{\cal F}(\boldsymbol{Q}) = \frac{1}{16}\int_{0}^{\infty}P^{3}dP\frac{12a^{4}+Q^{2}(2P^{2}Q^{2}-5a^{2})(a+2P^{2})}{a^{3}(a^{2}-P^{2}Q^{2})^{5/2}}
\end{equation} 
Finally, we integrate over momentum; using substitution $Q = 2 \sinh \theta$, the integral yields:
\begin{equation}
{\cal F}(Q=2\sinh\theta)=\frac{1}{64\cosh^{2}\theta}\left(1+3\frac{2\theta}{\sinh2\theta}\right),
\end{equation}
with the following asymptotic behavior:
\begin{equation}
\label{eq:AsymptoticF2D}
{\cal F}(Q) \approx \frac{1}{16}\cdot\begin{cases}
1, & Q\ll 1\\
1/Q^2, & Q\gg1
\end{cases},\quad (2D)
\end{equation}
\paragraph{3D case}
We use the same substitution $a = 1 + P^2 + Q^2/4$, and angular averaging yields:
\begin{multline}
{\cal F}(\boldsymbol{Q}) = \frac{1}{12\pi}\int_{0}^{\infty}dP\frac{P}{a^{3}}\Big[\frac{a+2P^{2}}{Q}{\rm arctanh}\frac{PQ}{a}-\\
-\frac{aP(a^{3}-4a^{2}P^{2}+2P^{4}Q^{2})}{(a^{2}-P^{2}Q^{2})^{2}}\Big].
\end{multline}
Again, using the same substitution $Q = 2 \sinh \theta$, the integral yields:
\begin{equation}
{\cal F}(Q=2\sinh\theta)=\frac{1}{192\cosh^{2}\theta}\left(2+\frac{1}{\cosh^{2}(\theta/2)}\right),
\end{equation}
with the following asymptotic behavior:
\begin{equation}
\label{eq:AsymptoticF3D}
{\cal F}(Q) \approx \frac{1}{192}\cdot\begin{cases}
3, & Q\ll1\\
8 / Q^2, & Q\gg1
\end{cases},\quad (3D)
\end{equation}

\end{document}